\documentclass{aa}

\usepackage[T1]{fontenc}
\usepackage{graphicx}
\usepackage{txfonts}
\usepackage{upgreek}
\usepackage{gensymb}
\usepackage{booktabs}
\usepackage[colorlinks=true,linkcolor=blue,citecolor=blue,urlcolor=blue]{hyperref}

\newcommand{\jwst}{\emph{JWST}\xspace}
\newcommand{\ha}{H\(\alpha\)\xspace}
\newcommand{\hb}{H\(\beta\)\xspace}
\newcommand{\pab}{Pa\(\beta\)\xspace}
\newcommand{\sirocco}{\textsc{Sirocco}\xspace}
\newcommand{\cloudy}{\textsc{Cloudy}\xspace}
\newcommand{\hei}{\ion{He}{i}\xspace}
\newcommand{\feii}{\ion{Fe}{ii}\xspace}
\newcommand{\feiin}{\ion{Fe}{ii}}
\newcommand{\nai}{\ion{Na}{i}\xspace}

\newcommand{\nii}{[\ion{N}{ii}]\xspace}
\newcommand{\caii}{[\ion{Ca}{ii}]\xspace}
\newcommand{\micron}{$\upmu$m\xspace}
\begin{document}
\title{Inside the cocoon: A comprehensive explanation of the spectra of little red dots}
\titlerunning{Inside the cocoon}

\author{A.~Sneppen\inst{1,2,3,4}\thanks{Corresponding author: \email{a.sneppen@gmail.com}}
\and D.~Watson\inst{1,2}
\and J.~H.~Matthews\inst{5}
\and G.~Nikopoulos\inst{1,2}
\and N.~Allen\inst{1,2}
\and G.~Brammer\inst{1,2}
\and R.~Damgaard\inst{1,2}
\and K.~E.~Heintz\inst{1,2}
\and C.~Knigge\inst{6}
\and K.~S.~Long\inst{7,8}
\and V.~Rusakov\inst{9}
\and S.~A.~Sim\inst{10}
\and J.~Witstok\inst{1,2}}
\authorrunning{Sneppen et al.}

\institute{
Cosmic Dawn Center (DAWN)
\and Niels Bohr Institute, University of Copenhagen, Jagtvej 155A, 2200 Copenhagen, Denmark
\and Institute of Astronomy, University of Cambridge, Madingley Road, Cambridge, CB3 0HA, UK 
\and Kavli Institute for Cosmology, University of Cambridge, Madingley Road, Cambridge, CB3 0HA, UK
\and Astrophysics, Department of Physics, University of Oxford, Oxford OX1 3RH, UK
\and School of Physics and Astronomy, University of Southampton, Southampton SO17 1BJ, UK
\and Space Telescope Science Institute, 3700 San Martin Drive, Baltimore, MD 21218, USA
\and Eureka Scientific Inc., 2542 Delmar Avenue, Suite 100, Oakland, CA 94602-3017, USA
\and Jodrell Bank Centre for Astrophysics, University of Manchester, Oxford Road, Manchester M13 9PL, UK
\and Astrophysics Research Centre, Queen's University Belfast, University Road, Belfast BT7 1NN, UK}

\date{Received \today; accepted ---}

\abstract{The James Webb Space Telescope (\jwst) has revealed a new population of compact galaxies, the `little red dots' (LRDs), whose nature remains unclear. We assemble grating spectra for 53 LRD-like objects and use self-consistent radiative-transfer calculations to show that a supermassive black hole embedded in a dense gas cocoon reproduces the spectra in detail, without requiring a stellar-like photosphere. The near-symmetric, electron-scattering-broadened emission line-shapes can be reproduced by cocoons with comparable inflowing and outflowing mass-fluxes. The same cocoon structure also simultaneously accounts for the continuum shapes, H and He line equivalent widths, and absorption features, and predicts metal-line absorption, X-ray weakness, and muted variability. We predict correlations should exist between Balmer break strength, Balmer line absorption, and scattering line width, and confirm their presence in the observed sample. Typical LRDs have black holes of order $10^{6}\,\mathrm{M}_\odot$ inside ionised cocoons of tens of $\mathrm{M}_\odot$, likely fed by a much larger cold-gas reservoir.}

\keywords{galaxies: high-redshift - (galaxies:) quasars: supermassive black holes - radiative transfer - line: profiles - scattering}

\maketitle

\begin{figure*}
\begin{center}
    \includegraphics[angle=0,width=1\textwidth]{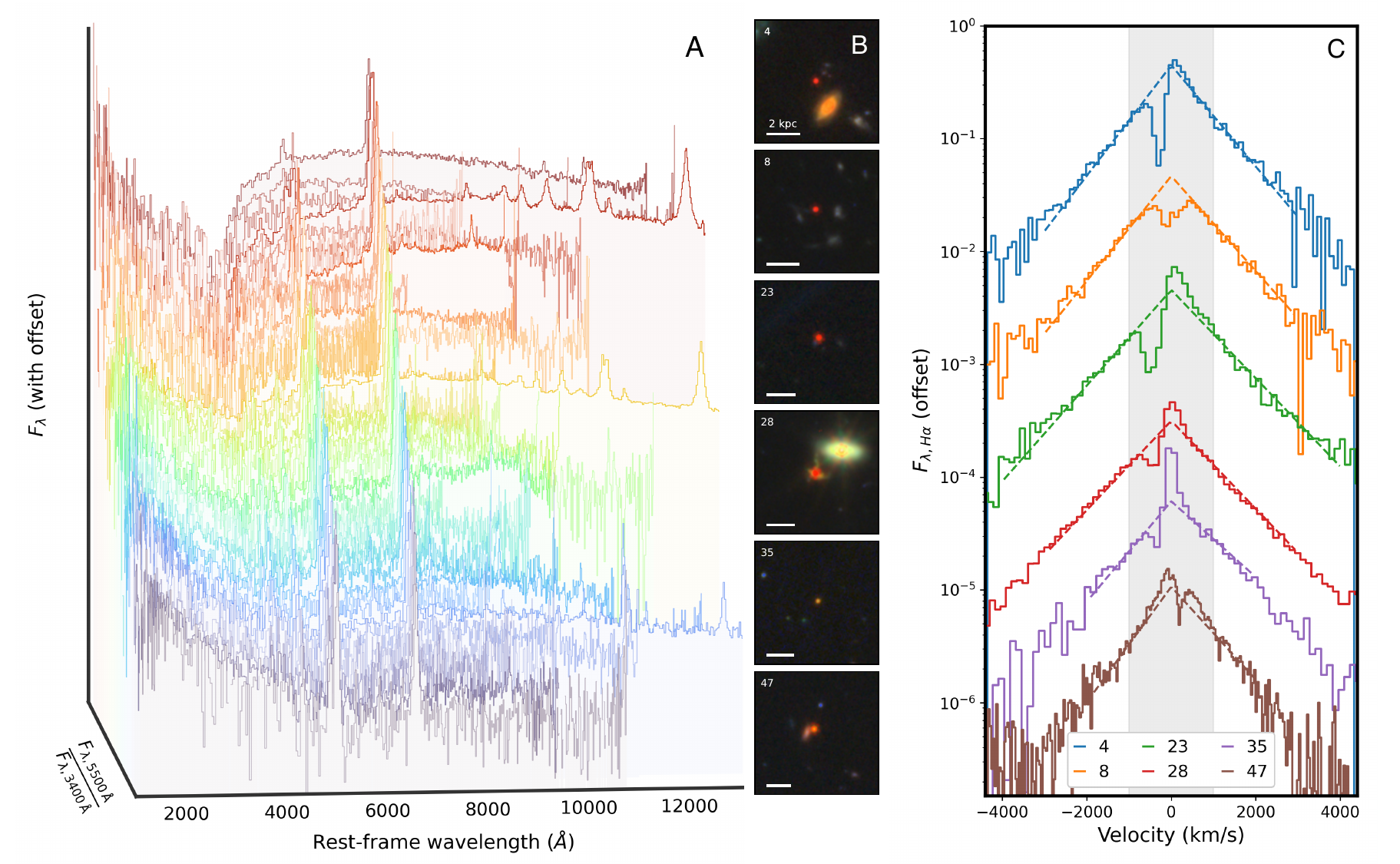}
\end{center}
\vspace{-0.6cm}
\caption{{Observed properties of the LRD sample ($2.2<z<7.2$).} \emph{A:} JWST\ NIRSpec/PRISM spectra show the characteristic LRD spectral shape with blue ultraviolet emission, red optical emission, and a change in slope near the Balmer break. \emph{B:} JWST\ images show that the sources are compact (images for the full sample are shown in Fig.~\ref{fig:images}). \emph{C:} H$\alpha$ emission-line profiles plotted on a logarithmic intensity scale. Exponential line wings appear as straight lines on this plot, while blue and red sides display similar exponential velocity scales.
}
\label{fig:observational_sequence}
\end{figure*}
\begin{figure}
    \begin{center}
    \includegraphics[angle=0,width=0.5\textwidth]{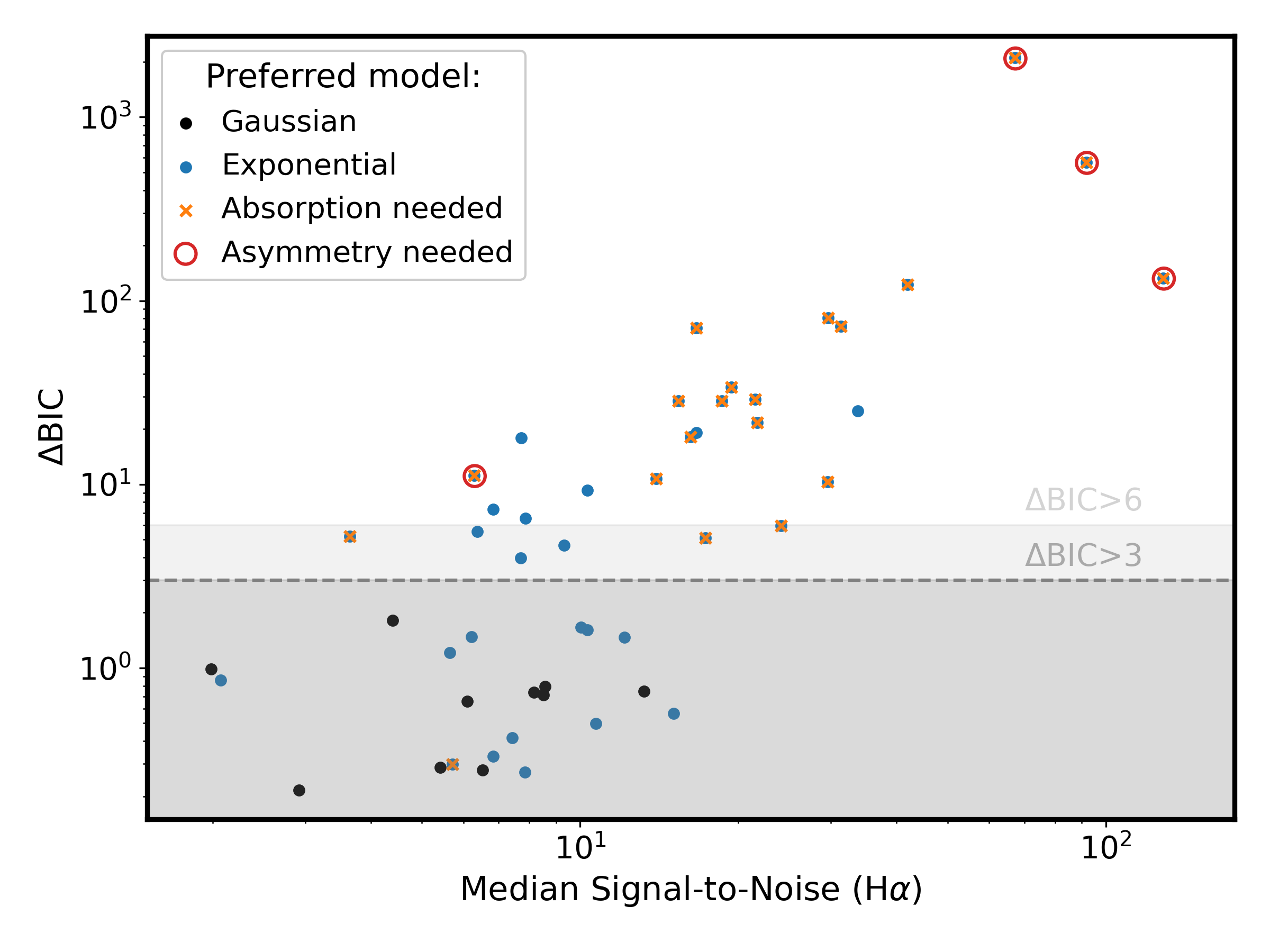}
    \end{center}
\vspace{-0.6cm}
\caption{{ \(\Delta\)BIC between \ha broad line fits}. We compare two-Gaussian models, representative of Doppler-motion-dominated broadening, with a single Gaussian convolved with an exponential profile, representative of electron-scattering-dominated broadening. Orange crosses and red circles indicate objects with evidence for absorption features and asymmetric exponential wings, respectively.}
\label{fig:deltaBIC}
\end{figure}

\begin{figure*}
\begin{center}
    \includegraphics[angle=0,width=0.98\textwidth]{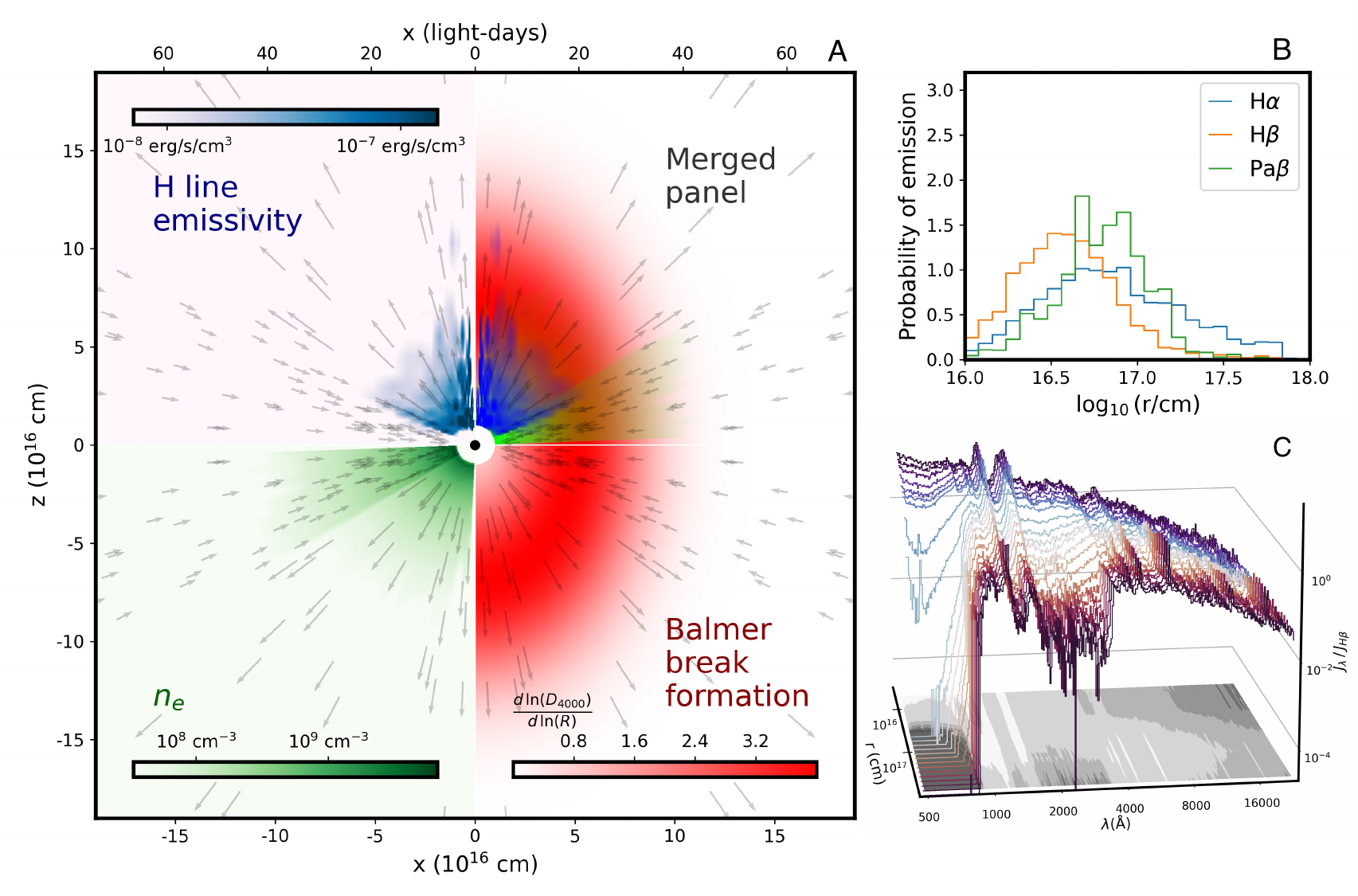}
\end{center}
\vspace{-0.3cm}
\caption{{Physical structure of an LRD-like object, showing the gas flows and the spectral-formation regions in the cocoon surrounding the central SMBH.} \emph{A}: Model is specified by a simple gas flow field: arrows show the gas velocity, with longer arrows indicating higher speeds, and their spatial distribution traces the gas density. The data require both inflow and outflow (see Fig.~\ref{fig:sphericity}). The quadrants show model outputs: colours indicate the region contributing to the hydrogen-line emissivity (blue), the electron density (green), and the Balmer break (red). 
\emph{B}: Radial origin of the escaping H$\alpha$, H$\beta$, and Pa$\beta$ emission, demonstrating that different hydrogen lines emerge from different radii.
\emph{C}: Local radiation field inside the cocoon. 
Hydrogen recombination lines form in the inner ionised gas, whereas the Balmer break arises further out.
}
\label{fig:cartoon_D}
\end{figure*}

\section{Introduction}\label{sec:introduction}

Little red dots (LRDs) are a population of luminous high-redshift sources recently identified by the \textit{James Webb} Space Telescope (JWST) \citep{Harikane2023,Matthee2024}, which exhibit an unusual combination of properties: i) they are spatially unresolved with JWST at least in the red part of the spectrum, implying sizes smaller than tens of parsecs \citep{deGraaff2025,Yanagisawa2026}; ii) they have a discontinuity in their spectrum at the rest-frame wavelength of $\sim$3646\,\text{\AA}, caused by hydrogen absorption and known as the Balmer break \citep{Setton2024}; iii) they display broad hydrogen and helium emission lines with widths of a few thousand km\,s\(^{-1}\) \citep{Matthee2024}; iv) with predominantly exponential line shapes \citep{Rusakov2025}; and v) they sometimes have relatively narrow red- and blueshifted H\,\textsc{i} Balmer and He\,\textsc{i} absorption \citep{Ji2025,Rusakov2025}. The LRDs belong to a more general set of objects with similar properties but with different colours, which are thought to have optical continua and broad lines dominated by ionised gas emission \citep{Rusakov2025,Naidu2025}. We refer to this more general set as `LRD-like' objects to subsume new classifications such as `little blue dots' \citep{Brazzini2026,Geris2026,Sneppen2026d}. So far, no model proposed has quantitatively and self-consistently explained all of their major features \citep[e.g.][]{Perez_Gonzalez2024,Inayoshi2025,Li2025,Bellovary2025}. 

The most favoured concept currently involves an accreting supermassive black hole (SMBH) due to their extreme compactness, broad H and He emission lines \citep{Matthee2024}, and extreme luminosities in small inferred volumes \citep{Rusakov2025}. However, beyond the difficulty of explaining the strong Balmer breaks and Balmer absorption lines and exponential line shapes, typical active galactic nucleus (AGN) indicators such as X-ray \citep{Yue2024,Maiolino_Chandra_2024}, radio \citep{Akins2024cosmoswebb}, and infrared excess \citep{Xiao2025,Perez_Gonzalez2024} emission are generally not detected, and only limited long-term variability has been found \citep{Zhang2025b,Burke2025}. In addition, SMBH masses based on single epoch virial methods are orders of magnitude `overmassive' \citep{Reines2015} relative to their host galaxies \citep{Pacucci2023,Maiolino_JADES_2024,Taylor2025}.

Modelling LRD-like objects as rapidly accreting SMBHs embedded in dense ionised gas \citep{Inayoshi2025,Liu2025,deGraaff2025b,Naidu2025,Rusakov2025} seems to be able to resolve some of these issues. First, the exponential form of the broad line wings indicates that the broadening is dominated by high column density electron scattering rather than Doppler motions \citep{Rusakov2025,Torralba2025,Kokorev2025}, implying smaller black hole masses \citep{Greene2005}. Second, such high columns could naturally give rise to the deep Balmer breaks and the Balmer and He\,\textsc{i} line absorption \citep{Matthee2024,Juodzbalis2024_rosetta,Naidu2025,Inayoshi2025,Liu2025,deGraaff2025}, deviations from the standard 'Case~B' hydrogen line ratios \citep{Nikopoulos2025,DEugenio2025b}, and suppression of both radio and X-ray emission \citep{Maiolino_Chandra_2024,Juodzbalis2024_rosetta,Rusakov2025}.

Here we present the first self-consistent, quantitative model of LRD-like objects to simultaneously reproduce the various observed properties. We compare a JWST sample of LRD-like objects, including several well-studied LRDs with extreme properties, to Monte Carlo ionisation and radiative-transfer simulations and show that the models reproduce the observed spectral properties in detail. Section~\ref{sec:sample_model} describes the sample, the line-profile fits ,and the radiative-transfer modelling framework. Section~\ref{sec:reproducing} compares the model spectra with the observed continua, line strengths, and line shapes. Section~\ref{sec:discussion} discusses the physical implications for the black-hole masses and Eddington ratios, the cocoon mass and feeding, as well as the predictions that can test this framework.

\begin{figure*}
\vspace{-0.1cm}
    \includegraphics[angle=0,width=\textwidth]{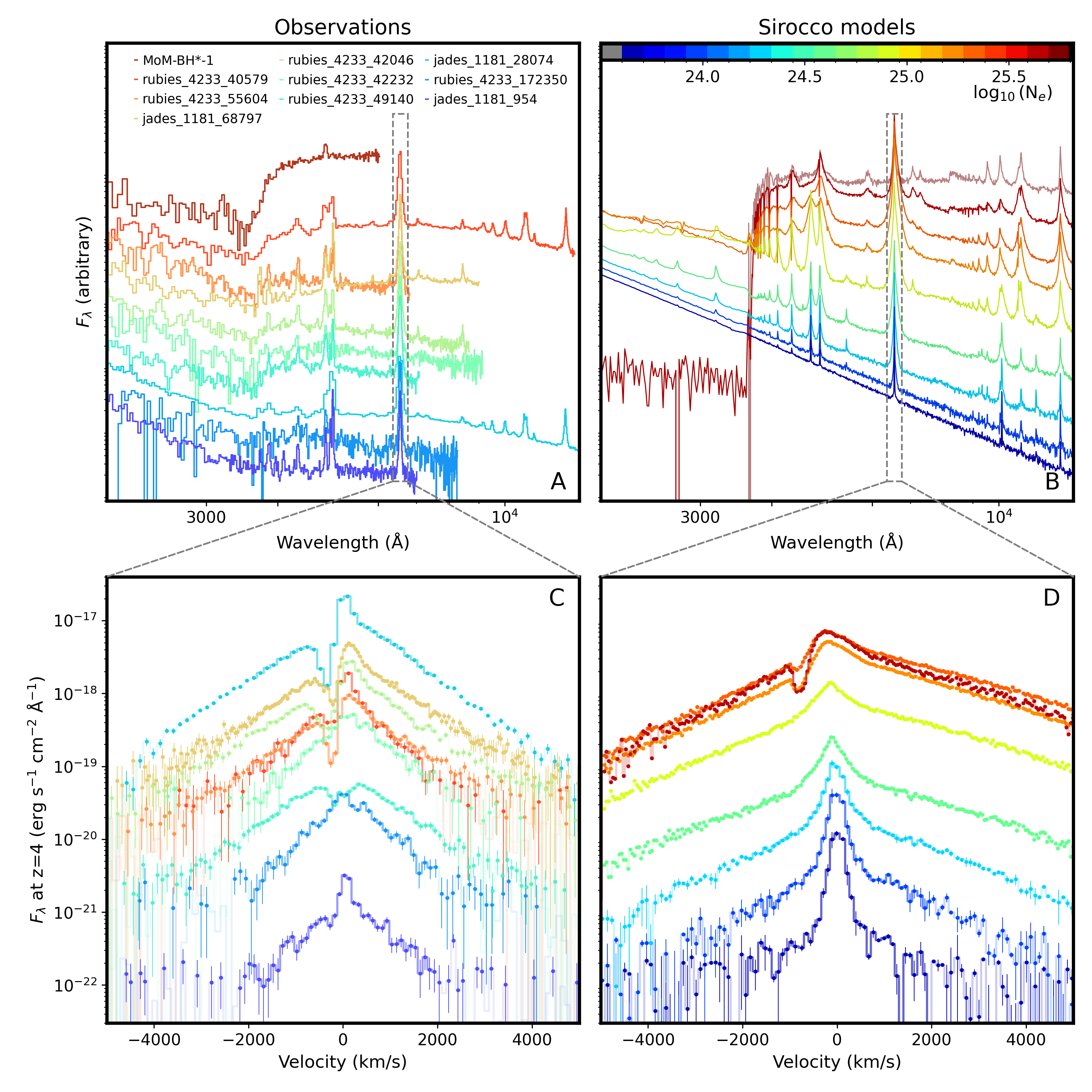}
\vspace{-0.7cm}
\caption{{Observed and model spectra of LRD-like objects ordered by Balmer break strength.} \emph{A:} Observed sources spanning a sequence of increasingly strong Balmer breaks. \emph{B:} \sirocco model spectra for a homologous wind with varying column density, keeping all other parameters constant. 
\textit{C}: \ha profiles for the observed sample. \emph{D:} \sirocco \ha profile scaled to the luminosity distance at $z=4$. In both the models and the observations, the systems with strongest Balmer break and most luminous \ha emission (i.e.\ the most massive cocoons) generally show broader wings and are more likely to exhibit Balmer absorption. }
\label{fig:exponential-sequence}
\label{fig:density-sequence}
\end{figure*}

\section{Observational sample and modelling framework}\label{sec:sample_model}
We select objects from the DAWN JWST\ Archive (DJA) with broad H$\alpha$, unresolved morphologies in the JWST\ F444W imaging band, and either a Balmer break or an inflection in the continuum slope near the Balmer-break wavelength \citep{DeGraaff2024_RUBIES,Heintz2025}; full selection details are given in Appendix~\ref{app:obs_sample}. These criteria yield 53 objects at \(z=2.2{-}7.2\). The sample does not include MoM-BH*-1 \citep{Naidu2025}, as it does not have a grating observation that covers \ha, but we still make quantitative predictions using \sirocco for such an extreme LRD. The colour, spatial, and broad-line structures of the sample are illustrated in Fig.~\ref{fig:observational_sequence}. 
We fit the \ha lines in the grating spectra. 
Objects have increasingly strong evidence of electron-scattering broadened lines as the S/N of the \ha lines increases, as quantified by the $\Delta$BIC (Fig.~\ref{fig:deltaBIC}), supporting the accuracy of the model. We tested exponential and Gaussian line shapes for these objects following \citet{Rusakov2025}, with a significance threshold of $\Delta\textrm{BIC}>6$ \citep{KassRaftery1995} and flat priors on all parameters. The instrumental line-spread function is forward-modelled using the NIRSpec dispersion curves of each grating, increased by a factor of 1.7 for compact sources. 
The broad line models compared were: the sum of two Gaussian broad lines ($\mathrm{gaus}$), a double exponential broad line plus a Gaussian un-scattered core ($\exp$), or an asymmetric exponential broad line plus a Gaussian un-scattered core ($\exp_{\mathrm{asym}}$) with and without absorption components. In each case we also included an additional narrow Gaussian component for \ha and narrow components associated with the \nii doublet. Twenty-three objects clearly favour exponential wing shapes, with 30 objects statistically indistinguishable, and no object favouring a two-Gaussian parameterization. All objects in the sample with prominent absorption features favour exponential wings, of which 18 are significant with $\Delta\textrm{BIC}>6$. 
The significance of these best-fitting models is assessed from  $\chi^2$ of the optimised model, computed with \textsc{dynesty} \citep{Speagle2020,sergey_koposov_2024_12537467}. The exponential line profiles are typically symmetric (see, for example, Fig.~\ref{fig:observational_sequence}C) and always close to symmetric with blue and red exponential widths consistent at $\lesssim$10\%. 

To interpret these spectra we use \sirocco \citep{long2002,Matthews2025}, an open-source Monte Carlo radiative-transfer code that self-consistently couples the cocoon's ionisation structure and radiation field \citep{Parkinson2025}. \sirocco is designed to treat optically thick, multi-dimensional flows including Compton scattering and relativistic effects -- properties not all available in the static and 1D codes used in previous LRD studies (see Sect.~\ref{subsec:rtcodes}). 

Our theoretical framework consists of an accreting SMBH whose radiation ionises a surrounding gas cocoon that is axisymmetric, with rotational Keplerian support around the SMBH and parameterized inflows and outflows. 
The diversity of our LRD objects can be modelled with accreting black holes with luminosities ($10^{43}{-}10^{45}\,\mathrm{erg\,s^{-1}}$) and a surrounding cocoon of densities \(10^{6}{-}10^{10}\,\mathrm{cm}^{-3}\), initialized between $R=10^{15}{-}10^{18}$\,cm with a falling radial density profile, fiducially $\rho \propto r^{-2}$. 
Observed Balmer and \hei absorption features have velocities in the range $-400$ to $+100\,\mathrm{km\,s}^{-1}$ \citep[e.g. best-fit model in our sample in App.~\ref{app:obs_sample},][]{Matthee2026,Yanagisawa2026b,Juodzbalis2026_abs}, indicating both inflowing and outflowing gas, while the relative symmetry of the electron-scattering wings supports a net gas flow close to zero, as discussed below. \sirocco permits a large variety of multi-dimensional velocity structures including convective and turbulent flows, but we fiducially only assume the observationally required inflows and outflows. 
The ionisation and excitation structure of LRDs emerges naturally from this framework, as shown in Fig.~\ref{fig:cartoon_D}: an inner, fully ionised region of high-column density electron-scattering gas extended beyond the Lyman-continuum Str\"omgren radius by Balmer-continuum photoionisation from $n=2$, and an outer partially ionised layer with a substantial $n=2$ hydrogen population that produces the Balmer absorption and the Balmer break.

Full model parameter descriptions and atomic data are given in \citet{Matthews2025}. The luminosity, density, geometry, and metallicity ranges explored are summarised in Appendix~\ref{app:sirocco_modelling}. We emphasise that the \sirocco spectra are not $\chi^2$-optimised fits to individual LRDs, but illustrative radiative-transfer calculations designed to test whether an optically thick scattering cocoon can reproduce the main observed spectral morphology and trends. A formal parameter search is currently computationally prohibitive, as each multi-dimensional Monte Carlo model requires of the order of 100 CPU hours. However, we quantify the comparison of \sirocco models and LRD observations using observable quantities such as line widths, equivalent widths, Balmer break strengths, Balmer decrements, absorption velocities, and line asymmetry.

\begin{figure*}
\begin{center}
    \includegraphics[angle=0,width=\textwidth]{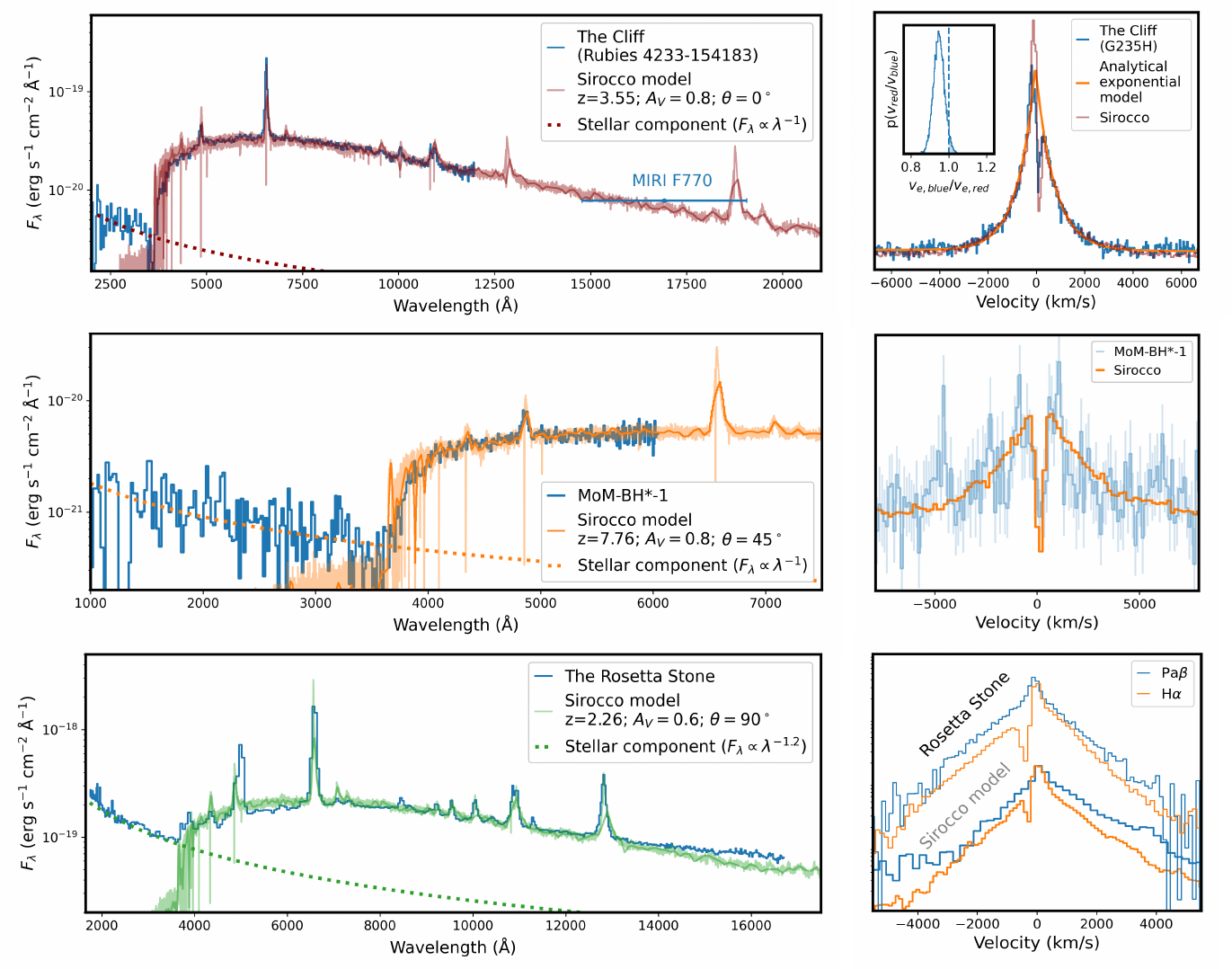}
\end{center}
\vspace{-0.5cm}
\caption{{Spectra of three extreme LRDs compared with dust-reddened \sirocco spectra.} The model spectra are matched to the observed bolometric luminosity, spectral shape, and shown for a viewing angle $\theta$ from the polar axis. \emph{A-C:} NIRSpec/PRISM spectra, with MIRI F770W photometry, of the Cliff \citep{deGraaff2025}, MoM-BH*-1 \citep{Naidu2025}, and the Rosetta Stone \citep{Juodzbalis2024_rosetta}. \emph{D-F:} Corresponding grating spectra showing \ha, \hb for MoM-BH*-1 and \pab for the Rosetta Stone. \sirocco models reproduce the detailed line morphologies of all three objects, including exponential wings, the distinct absorption components, and the \pab/\ha width ordering in the Rosetta Stone.}\label{fig:comparison_LRDs_Sirocco}
\end{figure*}

\begin{figure*}
\vspace{-0.1cm}
\begin{center}
    \includegraphics[angle=0,width=0.85\textwidth]{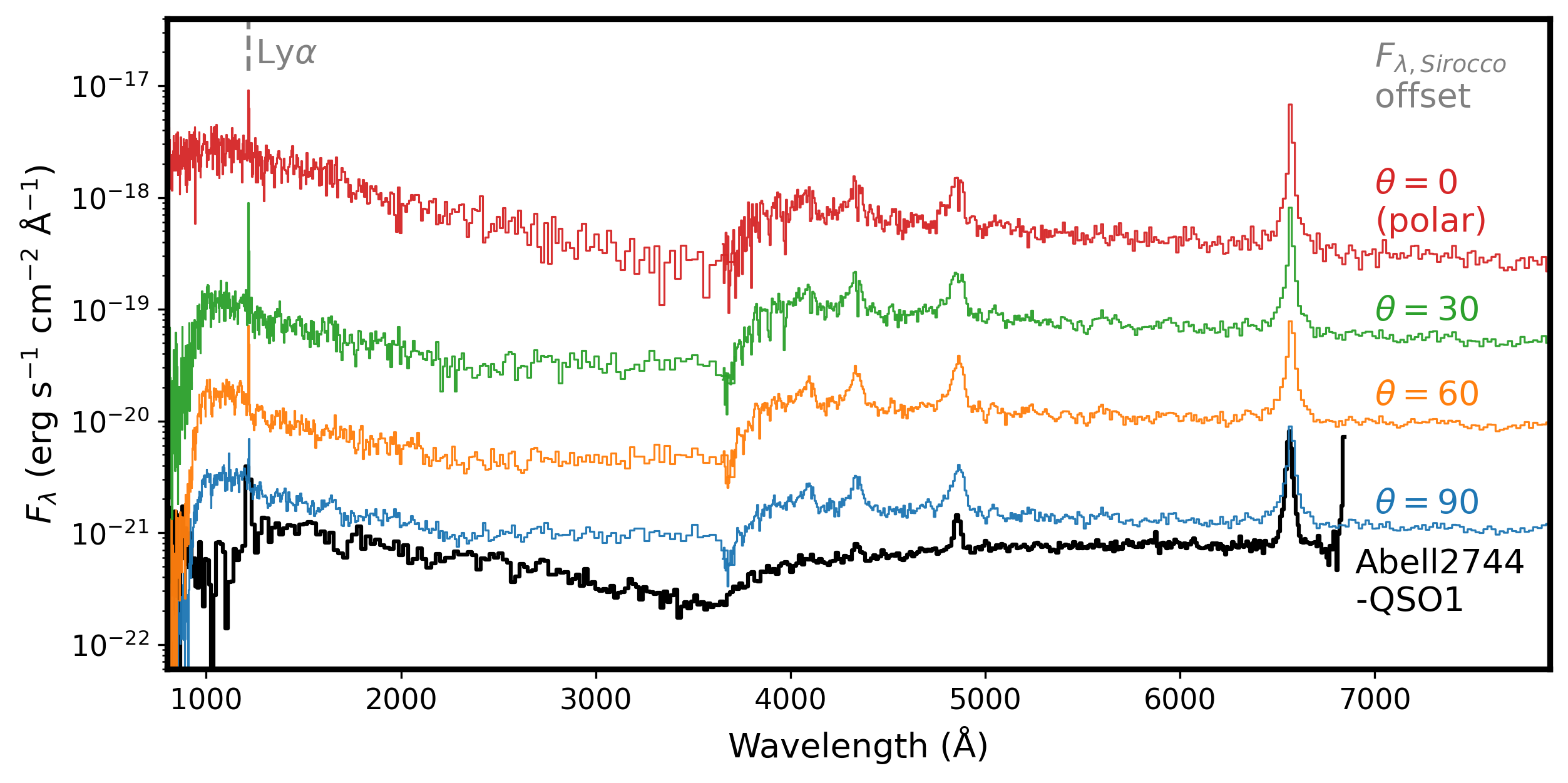}    
\end{center}
\vspace{-0.6cm}
\caption{{UV--optical spectrum of the LRD Abell2744-QSO1 compared with \sirocco model spectra at different viewing angles.} The model predicts a relatively featureless UV continuum, with prominent Ly-$\alpha$ emission and a smooth rollover across the Balmer break without requiring dust reddening ($A_V=0$).}\label{fig:UV}
\end{figure*}

\begin{figure}    
\vspace{-0.3cm}
\includegraphics[angle=0,width=0.95\linewidth]{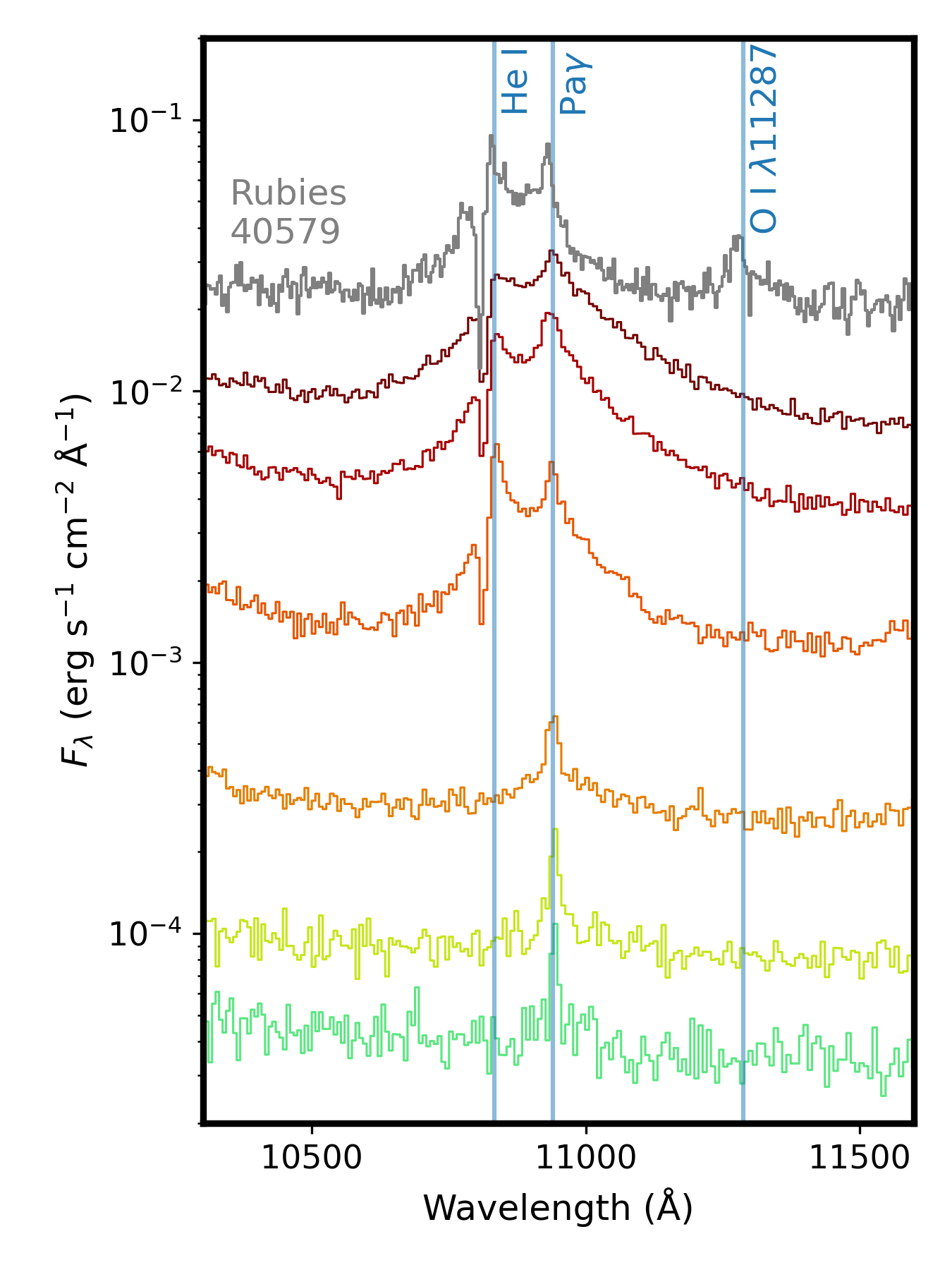}
\caption{{\sirocco model spectra around Pa$\gamma$ and \hei~$\lambda$10830.} The models correspond to the density sequence in Fig.~\ref{fig:density-sequence}. With increasing density, broad Pa$\gamma$ and \hei~$\lambda10830$ emission develops, accompanied by a superimposed blueshifted \hei~$\lambda10830$ absorption. RUBIES~40579, scaled to a common distance of 100\,pc, is shown for comparison and exhibits a similar morphology: broad emission with strong blueshifted absorption.}\label{fig:hei}
\end{figure}

\section{Reproducing LRD spectral properties}\label{sec:reproducing}

Fig.~\ref{fig:density-sequence} compares the observed sample with synthetic spectra from our models. A cocoon density sequence reproduces the hallmark LRD spectral features: a (non-stellar) Balmer break (arising from bound-free absorption, i.e.\ photoionisation, in dense, partially neutral gas \citep{Setton2024,Inayoshi2025}, broad hydrogen and helium recombination lines with exponential wings \citep{Rusakov2025,Torralba2025}, blueshifted Balmer absorption from the outflow with absorption sometimes deep enough to reach into the continuum, and a broad range of spectral shapes, with breaks close to the Balmer limit. In terms of  spatial appearance, they would be compact, with possibly more extended UV host-galaxy emission (see Fig.~\ref{fig:comparison_LRDs_Sirocco}). 

The models also simultaneously reproduce secondary features observed in LRDs: high-ionisation broad permitted emission lines from He with self-absorption superposed (see Fig.~\ref{fig:hei}) and low-ionisation metal lines from Ca\,\textsc{ii}, \nai, and \feii \citep[see Fig.~\ref{fig:baseline} and ][]{Lin2025,Ji2026b}, UV continua with Ly-$\alpha$ emission whose strength spans the observed range and limits \citep[see Fig.~\ref{fig:UV} and][]{Tang2026,Ji2026,Kageura2026}, deviations from Case~B (see Fig.~\ref{fig:density-sequence-caseB}), a spectral inflection near the Paschen limit \citep[see][]{Sneppen2026b}, cold gas regions with $T_e\lesssim4000$\,K \citep[see][]{Wang2026} and X-ray weakness \citep[see][]{Sneppen2026c}. The lower column-density part of the model sequence corresponds to an AGN population, similarly characterised by the cocoon reprocessing emission, exponential line-shapes and X-ray weakness but without strong absorption in the Balmer continuum and lines \citep{Sneppen2026d}, which may relate to the recently identified complementary population of `little blue dots' \citep{Brazzini2026,Geris2026}.

\subsection{Connecting continua and lines}
\begin{figure}
\begin{center}
    \includegraphics[angle=0,width=0.5\textwidth,viewport=10 2 448 408 ,clip=]{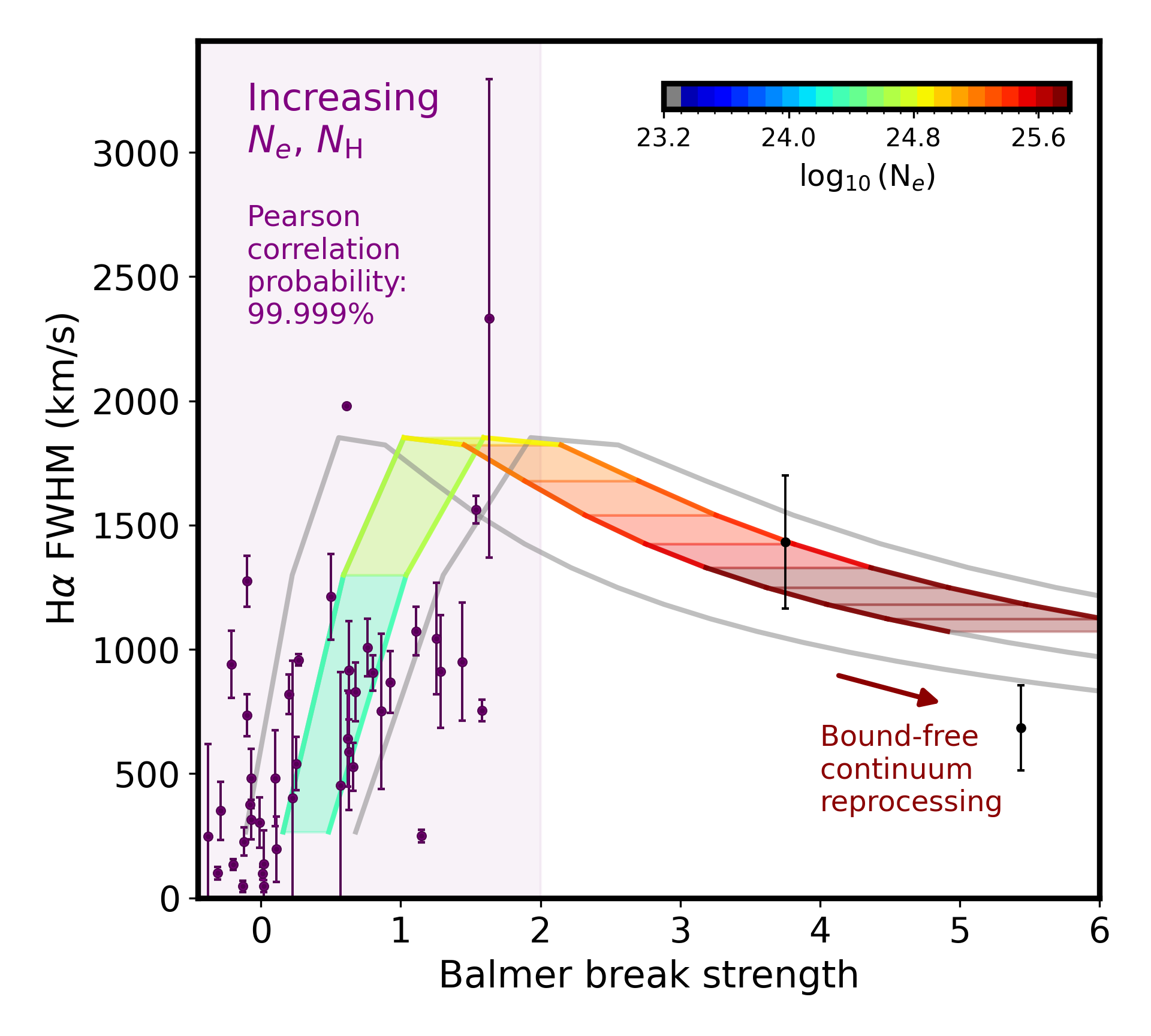}
\end{center}
\vspace{-0.7cm}
\caption{{\ha line-width versus Balmer break strength for the observed sample.} Points show the \ha full width at half maximum (FWHM), with error bars indicating 1$\sigma$ uncertainties. The coloured curves show an illustrative \sirocco density sequence reddened by a dust screen with $E(B-V)\in[0.4,0.8]$ ($E(B-V)=0$, $E(B-V)=1$ in grey), which shifts the models horizontally in Balmer break strength. Line width and Balmer break strength initially increase together as the electron-column density and $n=2$ hydrogen column rise with total gas column. At the highest columns, continuum absorption suppresses the multiply scattered wings, reducing the FWHM. The observed sources appear to follow this predicted pattern.}
\label{fig:corr_density}
\end{figure}

The various features probe different gas column densities. 
Strong electron-scattering wings require an electron column density $N_e\gtrsim10^{24}\,\mathrm{cm^{-2}}$, while the total hydrogen column density needed for absorption in Balmer lines and Balmer breaks is $N_H\gtrsim 10^{25}\,\mathrm{cm^{-2}}$ (for a hydrogen number density $n_H=10^{10}\,\mathrm{cm^{-3}}$). They are complementary diagnostics:
the magnitude of the Balmer break is sensitive to the gas density and the total column, the electron-scattering wings depend mainly on the ionised column density and electron temperature, while the strength of the recombination lines depends on the total ionised mass and the density. One example is the relation between \ha line width and Balmer break strength; our models predict a correlation up to moderate column densities (Fig.~\ref{fig:corr_density}). However, the line width saturates at $\sim3000\,$km\,s\(^{-1}\) because the most scattered photons are more likely to be absorbed before escaping. The observed sample follows this behaviour. No object in the sample is broader than $3000\,$km\,s\(^{-1}\) and the \ha width correlates with the Balmer break strength up to a break strength of about two (Pearson correlation probability of 99.999\%; see Fig.~\ref{fig:corr_density}). 

Consequently, objects with very extreme Balmer breaks should exhibit comparatively weaker and narrower recombination lines. We see this in The Cliff and MoM-BH*-1, both of which have exceptionally strong breaks but relatively weak and narrow lines (see Fig.~\ref{fig:comparison_LRDs_Sirocco}). In this regime, both the model spectra and the observed LRDs can show different exponential widths for different hydrogen lines (Fig.~\ref{fig:comparison_LRDs_Sirocco}F). The smaller width of \ha compared to \hb and \pab in the Rosetta Stone has been taken as evidence against electron scattering \citep{Juodzbalis2024_rosetta,Brazzini2025}; our models, however, reproduce this naturally, because electron scattering broadens all lines while wavelength-dependent bound-free absorption removes more photons from the wings of lines closer to a photoionisation edge. The non-monotonic ordering with wavelength further rules out dust extinction as the dominant cause. From the bound-free versus dust opacity comparison, we derive $M_{\rm dust}/M_{\rm gas}\ll 10^{-7}$ -- nearly five orders of magnitude below the Milky Way ratio (see App.~\ref{app:dust}).

A prediction of the model spectra is a tight correlation between the emission line strengths and the continua, since they are both created by recombination, that is, \ similar hydrogen-line equivalent widths (EWs) across objects. In other words, the Balmer line luminosities closely track the Paschen recombination continuum (the bound-free continuum due to recombination to hydrogen n=3 level; see Fig.~\ref{fig:Halpha_cont}). This trend has been reported in observed LRDs \citep{deGraaff2025b}, while it emerges naturally in our simulations because both the lines and the continuum are produced self-consistently by reprocessing of the same UV emission. The model suite shown in Fig.~\ref{fig:density-sequence} predicts large \ha equivalent widths, ${\rm EW(H\alpha)}=900\pm500\,\text{\AA}$, consistent with our observed LRD sample mean and scatter ${\rm EW(H\alpha)}=800\pm400\,\text{\AA}$. The \sirocco models predict that Balmer line-absorption strengthens with Balmer break strength ($B_{4200/3500}$, i.e. App.~\ref{app:obs_sample}); this trend is also present in the observational sample (Kendall p-value, $p = 0.4\%$). 

\subsection{Continuum interpretation}

The transmitted UV flux and the Balmer break produce the characteristic LRD-like SED. However, this configuration only occurs over a narrow density range ($\lesssim$0.5 dex) before the photons immediately bluewards of the Balmer break are very strongly suppressed (see Fig.~\ref{fig:density-sequence}). This is consistent with the observed LRDs, which are only compact in the reddest filters: with the AGN continuum suppressed bluewards of the break, stellar emission can more easily dominate the observed UV light. In most LRD-like objects, direct UV flux from the accretion region is not expected, explaining the absence of extreme ionisation signatures in their UV spectra \citep{Killi2024}. Across our simulation suite, the only UV line universally produced is Ly-$\alpha$ (see Fig.~\ref{fig:UV}), while other features become observable for hotter gas conditions (e.g. ${\rm C\,III]}\,\lambda\lambda1907,1909$, ${\rm O\,I}\,\lambda1302$, ${\rm C\,IV}\,\lambda\lambda1549$, and $1551$, alongside a plethora of UV metal absorption features; see Appendix~\ref{app:supp_text}). Broad, spatially unresolved Ly-$\alpha$ emission should be visible mainly in systems with little dust extinction and weak absorption by foreground neutral hydrogen, as appears to be the case in some recent LRD observations \citep{Ji2026,Tang2026}, albeit with a fairly rare incidence overall \citep{Kageura2026}.

Many proposed explanations for LRDs invoke a stellar-like atmosphere (quasi-stars and supermassive stars \citep{Begelman2025,Nandal2025} and a thermal or quasi-thermal continuum. Our simulations show that an LRD-like blackbody-adjacent continuum arises without any stellar surface: when fitted with modified blackbodies, the model spectra span the same shape range as the observed LRDs and reproduce the apparent Hayashi-like locus in temperature--luminosity space (\citealt{Kido2025,Inayoshi2025,deGraaff2025b}; Appendix~\ref{app:supp_text}). However, a simple blackbody framework does not predict a Paschen jump signature at $\sim$820\,nm, which is produced in our model spectra and observed \citep{Sneppen2026b}. Nor does it predict differences in the exponential widths of different hydrogen lines arising from wavelength-dependent bound-free opacity, which are predicted with \sirocco and observed \citep{Brazzini2025}. A strong correlation between the Paschen continuum and \ha emission, observed in the data \citep{deGraaff2025b} and predicted by \sirocco (see Fig.~\ref{fig:Halpha_cont}), naturally follows if both are produced by the same hydrogen and helium recombination, while an opaque, cold, and exterior photosphere would absorb or scatter \ha photons before they escape, suppressing \ha relative to the continuum, inconsistent with the data. The rest-frame optical and near-infrared continuum in LRDs is therefore almost certainly dominated by Paschen and Brackett recombination emission, not by a stellar atmosphere.

\subsection{UV emission in \sirocco}

While a significant fraction of the UV emission in LRDs -- and often the dominant contribution -- originates from the (star-forming) host galaxy \citep{Rinaldi2025,Baggen2026,PerezGonzalez2026}, a component of the UV emission appears to be intrinsic to the LRD phenomenon itself \citep{Sun2026}. In particular, the presence of spatially unresolved, broad Ly-$\alpha$ emission provides fairly direct evidence that UV photons can escape from the envelope and/or cocoon \citep{Furtak2024,Ji2026,Tang2026}. In our models, Ly-$\alpha$ is the only UV emission feature that is robustly present across the full simulation suite (see Fig.~\ref{fig:UV}). Ly-$\alpha$ photons, though more strongly attenuated by dust, are less strongly attenuated by bound-free continuum absorption than photons near $\lambda\sim3600$\,\text{\AA}, since the corresponding bound-free absorption opacity is $\sim$25 times smaller at Ly-$\alpha$ than at the Balmer break. With model Ly-$\alpha$/H$\alpha$ from about the Case B value at low densities to $\lesssim$1\% of it in the densest cocoons, the models are consistent with Ly-$\alpha$'s upper limits \citep{Kageura2026} and observations \citep{Ji2026,Tang2026}. 
Although our models predict broad Ly-$\alpha$ emission, we emphasise that the Ly-$\alpha$ line shapes produced may not be fully reliable in \sirocco due to the Sobolev approximation employed. Modelling the line profile requires a treatment of high-optical-depth resonant scattering together with photon-destruction channels, such as bound-free absorption processes, which can become important because resonant scattering greatly increases the photon path length. This is analogous to the coupling of electron-scattering and bound-free absorption opacity, which produces the line-dependent exponential widths of different permitted lines. 

Other UV spectral features present in a subset of the \sirocco simulations include ${\rm C\,III]}\,\lambda\lambda1907,1909$, ${\rm O\,I}\,\lambda1302$, ${\rm Mg~II}$, ${\rm Al~III}$, and ${\rm Fe~II}$, all of which may have observational counterparts in LRDs \citep{Akins2025_UV,Torralba2025,Tang2026,Kokorev2026}. These features are not necessarily broad in the simulations, although the width of ${\rm C\,III]}$ is model-dependent. By contrast, features only associated with the subset of models with hotter and more highly ionised cocoons, such as ${\rm C\,IV}\,\lambda\lambda1549,1551$, ${\rm He\,II}\,\lambda1640$, $\,\lambda3203$, and $\,\lambda4686$, may only have modest evidence of detections \citep{Labbe2024,wang2025missing}. This may point to the comparatively cold gas conditions in the observed gas, as suggested by independent lines of evidence \citep{Lin2025,Torralba2025}. Crucially, any detection of broad high-ionisation emission, such as ${\rm He\,II}\,\lambda4686$ \citep{wang2025missing}, necessitates some hard photons from the central ionising source, particularly as the emergent line strengths in \sirocco are strongly suppressed by absorption, as such photons propagate out from the highest optical depth regions of the cocoon.

Additionally, Fig.~\ref{fig:UV} illustrates the `V-shaped' SED shape and the smooth rollover shape in the optical continuum predicted by \sirocco. Here, accurate modelling of the Balmer break shape requires the inclusion of a high number of energy levels of hydrogen, with our modelling currently using up to $n=40$. A further softening of the rollover can occur due to higher electron-scattering optical depth, mild turbulent velocities, or a cooler recombination continuum. The UV spectral slope is directionally dependent, because of sight-line variations in the balance of bound-free emission (i.e. Balmer continuum emission) and bound-free absorption (i.e. Balmer continuum absorption), and potentially from variable neutral gas columns and, by extension, dust columns.

\begin{figure}
\vspace{-0.1cm}
\includegraphics[angle=0,width=\linewidth]{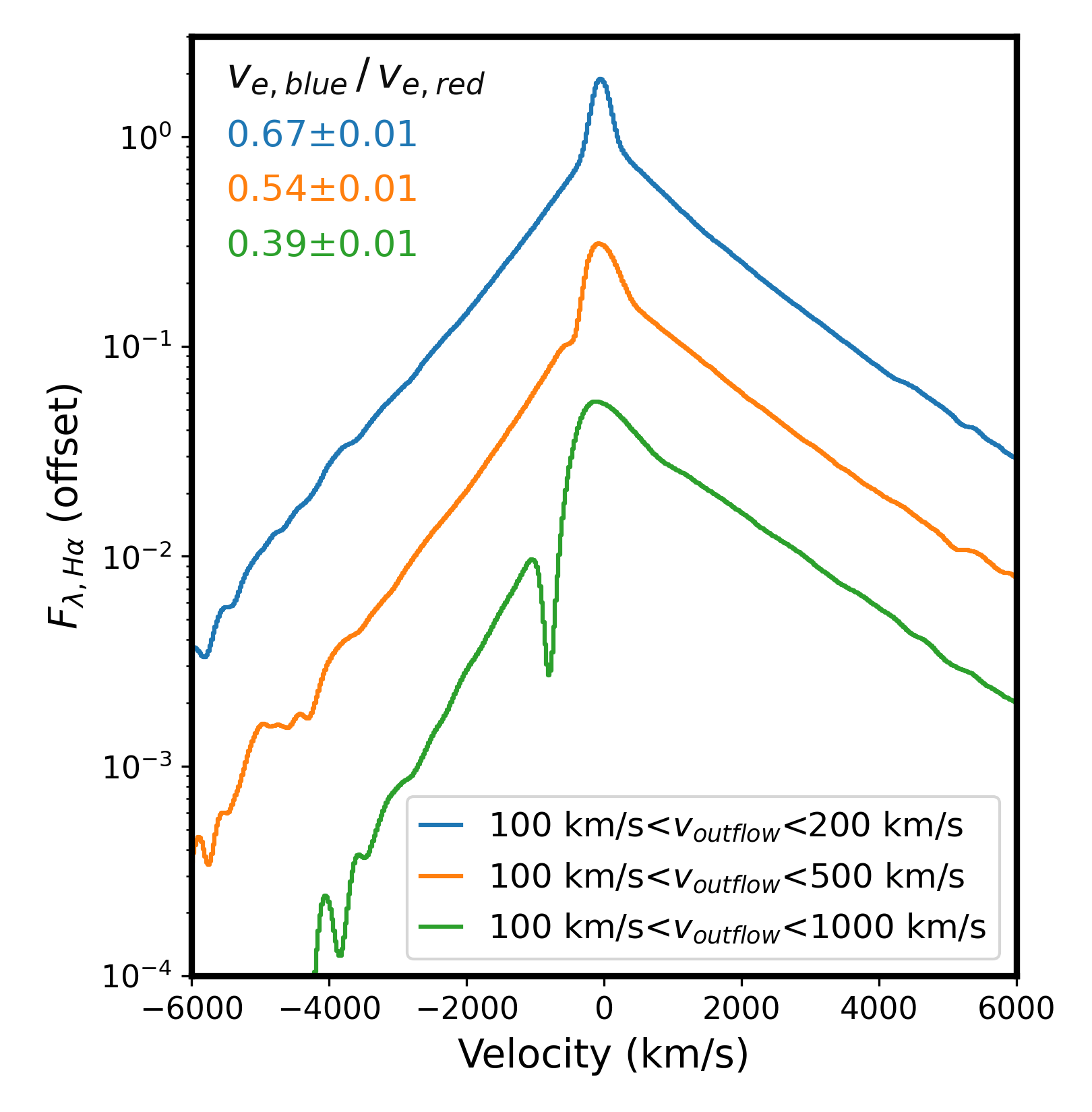}
\vspace{-0.6cm}
\caption{{Line shapes for a spherical electron-scattering medium at various wind velocities.} At velocities of $\sim500 \mathrm{\,km\,s}^{-1}$, the asymmetry between the blue and red exponential wings is dramatic. To self-consistently produce both symmetric scattering and the absorption and/or emission profiles, we must go beyond spherical models.}\label{fig:wind_velocity}
\end{figure}

\begin{figure}
\vspace{-0.25cm}
\includegraphics[angle=0,width=0.49\textwidth]{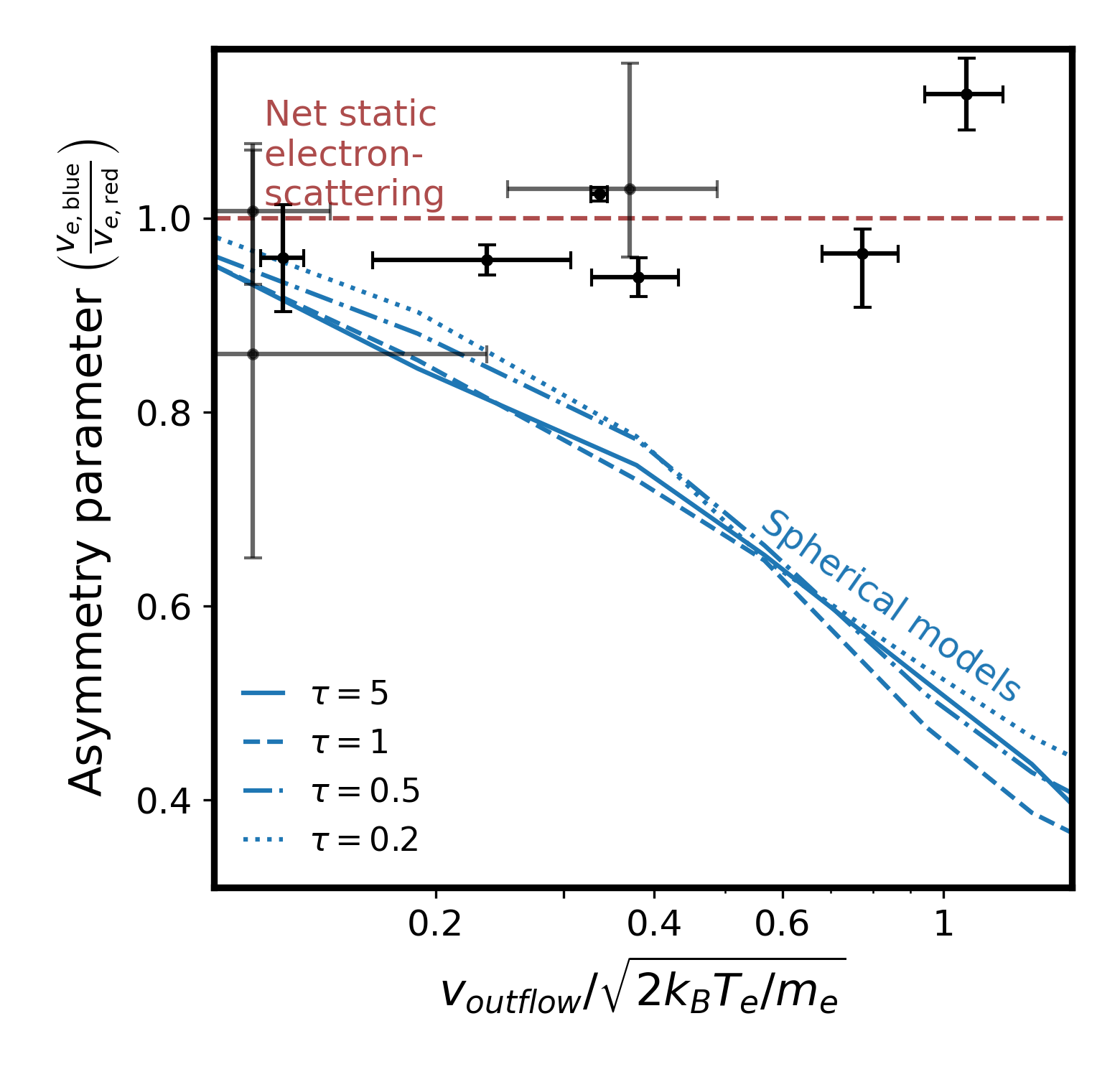}
\caption{{\ha wing asymmetry as a function of outflow velocity.}
Blue curves show the asymmetry predicted for spherical kinematics, while 1$\sigma$ error bars indicate sample objects with blueshifted absorption. Objects are often strongly discrepant with a spherical outflow, implying that both inflow and outflow are required. Asymmetry is defined as the ratio of the red- to blue-side exponential slopes, and the outflow velocity is measured from \ha absorption in units of the thermal electron velocity. The models do not predict perfectly symmetric profiles, and indeed, several objects show small, but measurable asymmetry.}\label{fig:sphericity}
\end{figure}

\begin{figure*}
\vspace{-0.3cm}
\begin{center}
     \includegraphics[angle=0,width=\textwidth]{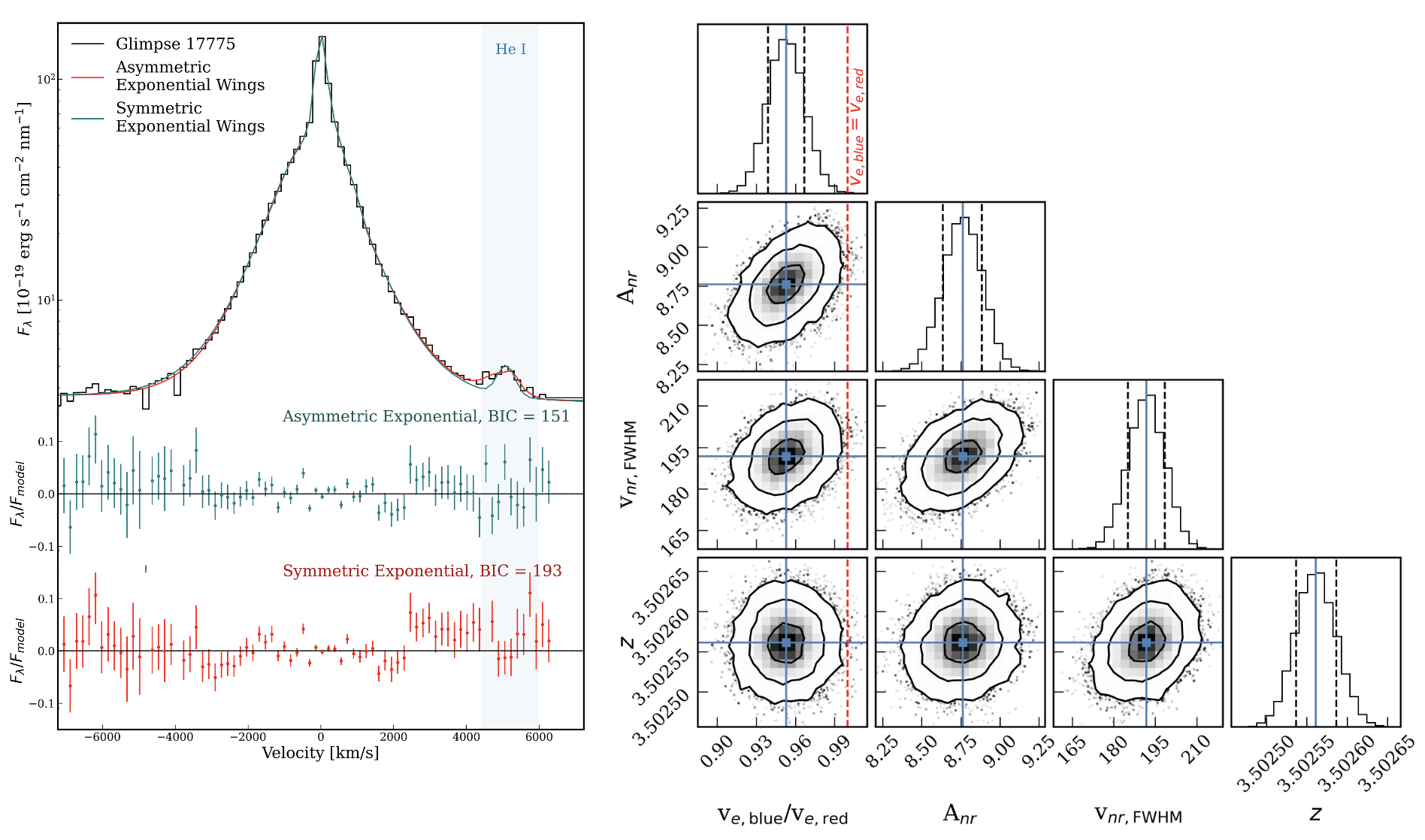}
 \end{center}
 \vspace{-0.6cm}
 \caption{{Example fit, residual structure, and posterior constraints of (a) symmetric electron-scattering wings for the highest S/N LRD \ha line, Glimpse 17775 \citep{Kokorev2025}.} While the symmetric electron-scattering model is superior to the double Gaussian broad line ($\Delta \mathrm{BIC}>$100), the asymmetric electron-scattering-wing model is further superior to the symmetric model ($\Delta \mathrm{BIC}$=42). The direction of asymmetry is as predicted for a net outflow. The posterior distribution is well behaved, with a clearly defined minimum. The observed line shape helps separate the effects of the asymmetry parameter, $v_{\rm e,blue}/v_{\rm e,red}$, from those of the other parameters, including the narrow line parameters, where amplitudes are in units of $10^{-19}\,\mathrm{erg\,s^{-1}\,cm^{-2}\,\text{\AA}^{-1}}$, and velocities in units of $\mathrm{km\,s^{-1}}$. 
}
 \label{fig:Glimpse}
\end{figure*}

\subsection{Hydrogen line-shapes -- non-spherical kinematics}
The observed line profiles show that most LRD-like sources cannot have a simple spherical velocity field. A significant fraction show absorption lines shifted by a few hundred km\,s\(^{-1}\) \citep{Juodzbalis2025,Rusakov2025}, which is comparable to the thermal velocities of the electron gas at about 10\,000\,K. For a spherical outflow, this would result in highly asymmetric scattering wings (see Fig.~\ref{fig:wind_velocity} or App~\ref{app:asymmetry}). 
The systematic redwards skew of the line centre arises from the net divergent outflow: when photons scatter into non-radial directions, the bulk velocity component away from the source imparts a net redshift \citep{Huang2018}. Thus, outflows tilt the profile redwards (i.e.\ a steeper blue exponential), whereas inflows tilt it bluewards (i.e.\ a steeper red exponential).
This effect is seen in the spectra of symbiotic stars \citep{Seker2012}, Wolf--Rayet stars \citep{Hillier1991}, tidal disruption events \citep{Roth2018,Parkinson2022}, luminous blue variables \citep{Peng2025}, luminous red novae \citep{Sneppen2026_LRNe}, and Type~IIn supernovae \citep{Roming2012,Humphreys2012}.\footnote{Symmetric exponentials have also been observed for Type~IIn SNe, but these are associated with narrow intrinsic lines in low-velocity winds; e.g.\ SN~1998S and SN~2010jl \citep{Chugai2001,Fransson2014}.} Parameterizing the blue and red wings with separate characteristic velocities
\[
F(v)=
\begin{cases}
F_0\,\exp\!\big(-{|v|}/{v_{\mathrm{e,blue}}}\big), & v<0,\\
F_0\,\exp\!\big(-{|v|}/{v_{\mathrm{e,red}}}\big),  & v>0,
\end{cases}
\]
we see $v_{\mathrm{e,blue}}/v_{\mathrm{e,red}}\simeq 1$ to within 10\% in our sample, with uncertainties for individual objects at the 5\% level. In one-dimensional models, this near-equality would imply that the bulk flow speed is small compared to the electron thermal speed $v_{\rm bulk}\lesssim0.1 \sqrt{2kT_e/m_e}=50\mathrm{\,km\,s}^{-1} \,(T/10^{4}\,\mathrm{K})^{1/2}$ (see Fig.~\ref{fig:sphericity}). The simplest explanation of both symmetric exponentials and high observed absorption line velocities requires a departure from spherically symmetric kinematics.

In Fig.~\ref{fig:sphericity} we compare the line symmetry of the objects with blueshifted absorption lines to the results of spherical outflow models as a function of outflow velocity, showing that the observed line symmetry is inconsistent with a spherical outflow at the observed absorption velocities. The coexistence of inflow and outflow indicates accretion from outside the ionised cocoon and a feedback-driven outflow. Different observed line profiles can then arise because different sight-lines pass through different parts of the flow: equatorial inflows in some directions and polar outflows in others (see Fig.~\ref{fig:viewing}).

\begin{figure*}
\begin{center}
    \includegraphics[angle=0,width=0.95\textwidth,viewport=50 45 1000 385 ,clip=]{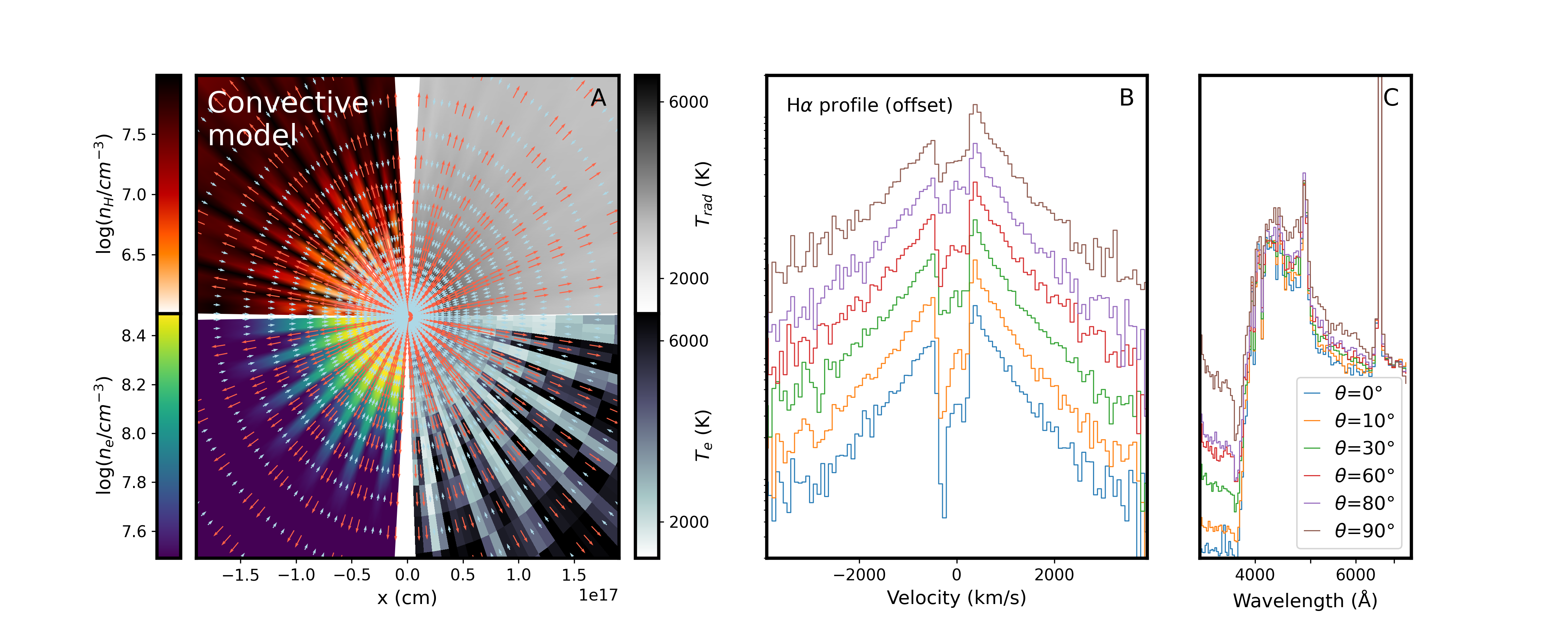}
    \includegraphics[angle=0,width=0.95\textwidth,viewport=50 45 1000 385 ,clip=]{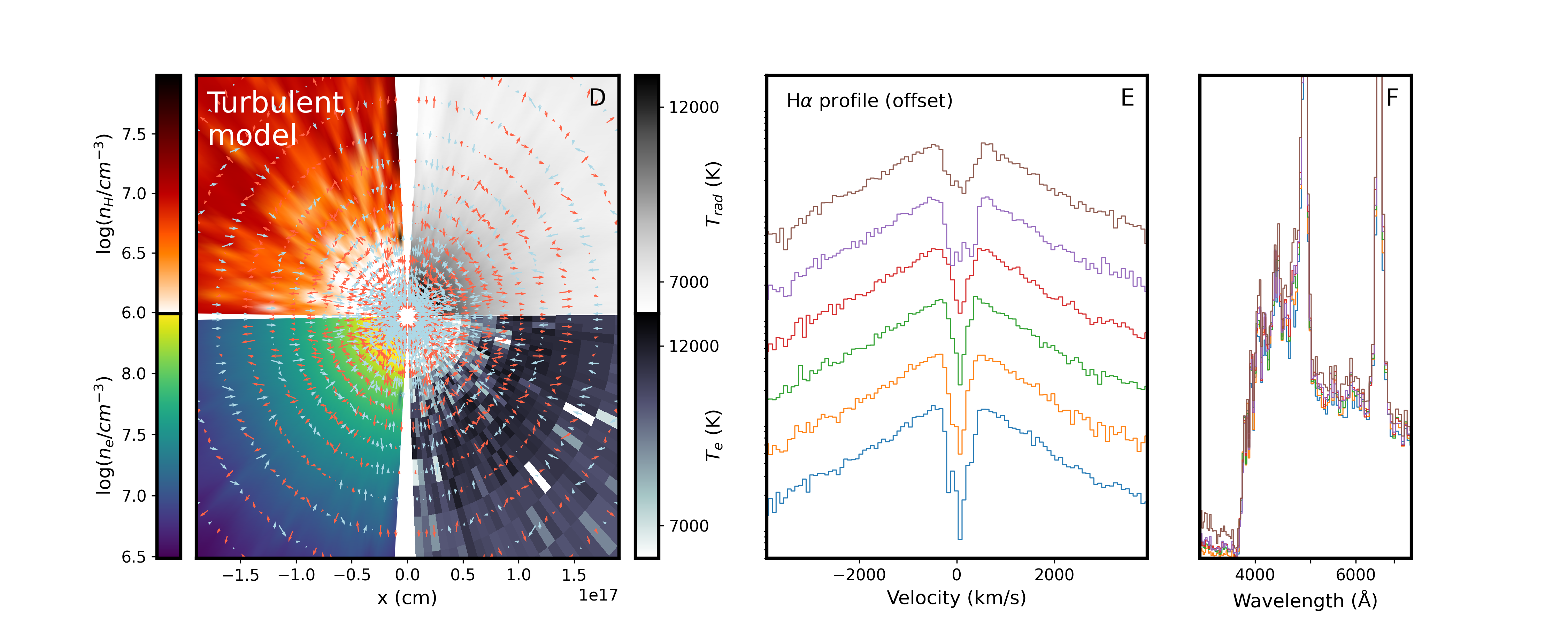}
    \includegraphics[angle=0,width=0.95\textwidth,viewport=50 14 1000 385 ,clip=]{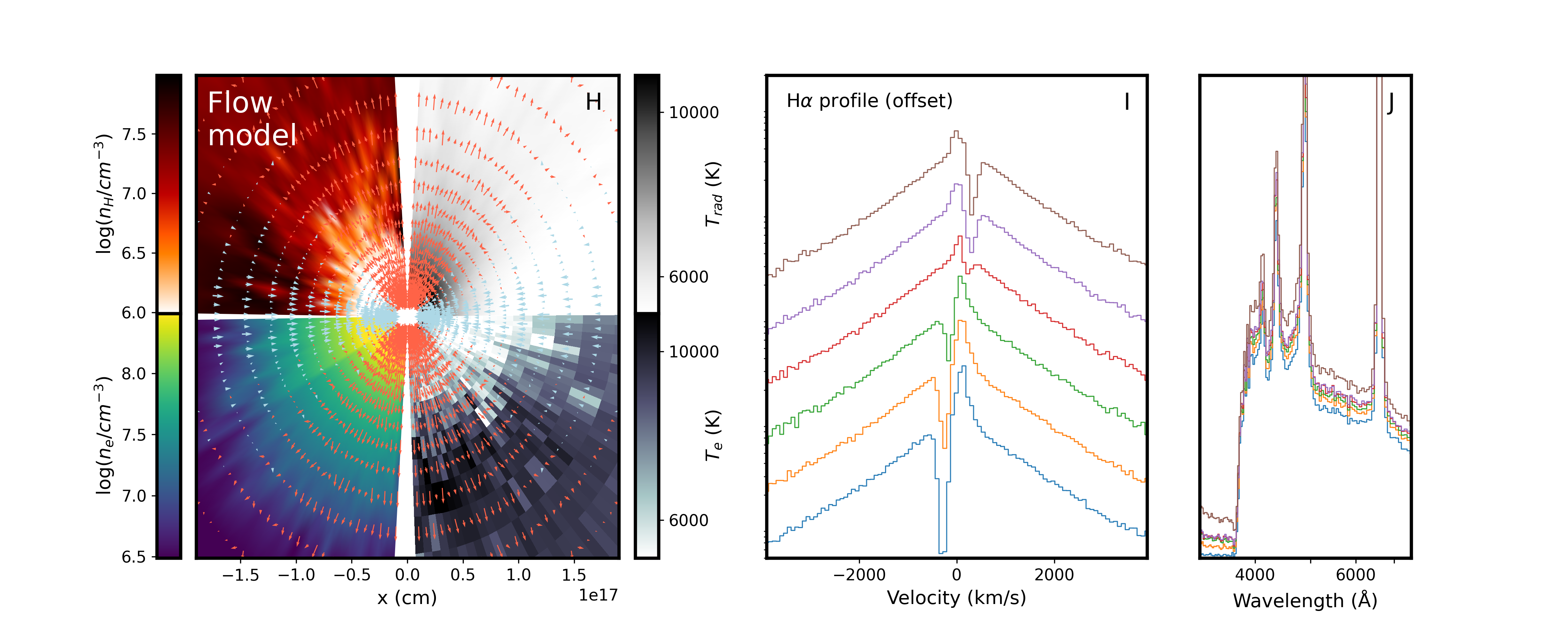}
\end{center}
\caption{{\sirocco LRD models with kinematics dominated by large-scale radial convective cells (\emph{top}), turbulent motion (\emph{middle}) and inflow- and/or outflow-dominated model (\emph{bottom}).} \emph{Left}: Neutral hydrogen density, radiation temperature, electron density, and temperature with a velocity field indicated by arrows (red being outflow and blue indicating inflow). \emph{Centre:} \sirocco model \ha line profile from various viewing angles (polar is $\theta=0$). \emph{Right:} \sirocco model continuum SED. A convective model shows a complex absorption profile, turbulent models show centred absorption, while inflow and/or outflows produce both blue- and redshifted absorption features. 
} \label{fig:convective_turb}
\end{figure*}

However, a perfect inflow and/or outflow balance would be fine-tuned, and on careful inspection, several high S/N objects in the sample show a small but significant redwards asymmetry in their broad lines (Figs.~\ref{fig:sphericity} and \ref{fig:Glimpse}), consistent with a denser, slower inflow than outflow (Appendix~\ref{app:asymmetry}). The observations necessitate outflows and inflows, but more broadly \sirocco can also accommodate diverse input velocity structures, such as convective or turbulent gas (see Fig.~\ref{fig:convective_turb}) or a prescribed structure of a numerical radiation-hydrodynamical model.  

\subsection{\sirocco models in the context of other radiative-transfer codes}\label{subsec:rtcodes}

In terms of modelling, \sirocco provides the first (and currently only) simultaneous reproduction of LRD lines and continua, including line-shapes, relative widths, and equivalent widths. Further, it provides the needed treatment of flows and departures from spherical symmetry.

However, recent radiative transfer work has explored a range of physical interpretations for the spectra of LRDs, providing useful insights and context for our modelling. Firstly, important insights have been gained from optically thin and 1D codes like \cloudy, including the non-stellar origin of Balmer breaks \citep{Inayoshi2025,Ji2025,Naidu2025,deGraaff2025,Taylor2025,Pacucci2026}, on the hard ionising photons \citep{wang2025missing}, as well as on the density and temperature of the ionised gas with [\feiin] emission \citep{Torralba2025,Lin2025}. Such slab calculations include line emission, but not detailed velocity-dependent line-profile radiative transfer. However, the details of resonance, Raman, and electron scattering for hydrogen line formation in such dense gas under the assumption of a static spherical gas distribution have been assessed \citep{Chang2025}. The latter work found that at high electron number density ($n_e>10^8\,\mathrm{cm}^{-3}$), \ha and \hb can have different emission regions. This implies different optical depths for electron scattering, resulting in a difference in the strength and width of the broad component around the hydrogen emission lines, which provides an alternate explanation for the variable exponential width observed \citep{Brazzini2025}. Previous semi-analytic \citep{Laor2006,Rusakov2025} and radiative transfer studies \citep{Rusakov2025,Chang2025} assumed a static electron gas for AGNs and LRDs, but by including bulk velocities we can test for absorption kinematics and diagnose scattering-wing asymmetries induced by bulk motion.

A complementary class of models has focused on the continuum emission. \citet{Liu2025,Liu2026} modelled LRDs as super-Eddington accreting black holes in dense, optically thick envelopes using semi-analytical atmosphere models and radiation transport. Such models can account for the red optical continua and strong, non-stellar Balmer breaks of LRDs. Indeed, a spectral library using stellar atmosphere codes can provide similar roughly thermal-looking spectral continua \citep{Liu2026}, although without reproducing the modelling of line emission in this regime of large stellar atmosphere-like columns ($N_\mathrm{H}\gtrsim10^{27}\,{\rm cm^{-2}}$), where the radiation field becomes cool and thermalised. Within \sirocco such cool input emission can also be modelled, but ultimately such cool models cannot simultaneously produce strong recombination emission, including from high-ionisation species such as He, and therefore require separate (magnetic) coronae to produce this recombination emission \citep{Gentile2026}. Alternatively, hotter input stellar atmospheres, $\gtrsim10\,000\,$K, can reproduce electron-scattered broad H and He line emission and a red continuum if subsequently reddened \citep{Martins2026}. However, they do not simultaneously reproduce the UV emission, the Balmer break, or the spectral inflection near the Balmer limit, and they re-radiate most of the input luminosity through dust emission, in tension with infrared observations \citep{PerezGonzalez2026}.

\section{Physical implications of dense cocoons}\label{sec:discussion}

Using our quantitative framework for modelling LRD-like AGNs, we can determine the cocoon's mass, electron temperature, column density, dust content, continuum absorption opacity, and kinematics. The model also enables the SMBH mass to be estimated without recourse to low-redshift scaling relations. The recombination luminosity and gas column set the characteristic emission radius, while the width of the un-scattered broad-line core gives the orbital velocity at that radius. This also allows us to measure the Eddington ratio, $L/L_{\rm edd}$, where $L_{\rm edd}$ is the luminosity at which the radiation pressure balances gravity.

Accounting for electron-scattering broadening, we find the intrinsic width of the broad lines in our sample. Standard AGN scaling relations \citep{Reines2013} imply SMBH masses \(\log_{10}(M_{\rm BH}/M_{\odot})=5.0{-}7.0\). By contrast, \sirocco does not rely on a luminosity--size scaling: the recombination luminosity and measured column density set the cocoon scale, while the un-scattered broad-line width gives the orbital velocity at that scale. Matching the un-scattered line components to our models yields \(\log_{10}(M_{\rm BH}/M_{\odot})=6.1\pm0.7\), where the value and uncertainty denote the median and 16th--84th percentile range. For the observed luminosities, these masses imply Eddington ratios of $\log_{10}(L/L_{\rm edd})=0.0\pm0.6$. Most objects in the sample are therefore consistent with accretion near the Eddington limit. 
The mass of gas in the ionised cocoon is small: $1-100$\,M$_\odot$. Typical electron temperatures vary across the cocoon from about 4000\,K to about 10\,000\,K. The hydrogen column densities range from about $10^{24}$\,cm$^{-2}$ to a few $10^{25}$\,cm$^{-2}$. As mentioned above, the dust content in the ionised cocoon is extremely low, essentially negligible. The velocities are $\lesssim100$\,km\,s$^{-1}$ in the inflow, up to 400\,km\,s$^{-1}$ in the outflow, and these two components contain comparable mass fluxes. 

With such small ionised-cocoon masses and high Eddington ratios, the depletion timescale is short, of the  order of a few thousand years, depending on the accretion efficiency. Therefore, a much larger cold-gas reservoir is needed to feed the cocoon if the LRD phenomenon is to survive more than a few millennia; a straightforward extrapolation of the cocoon structure to giant molecular cloud scales would yield a total reservoir mass $\sim 10^{6}\,M_\odot$ (Appendix~\ref{app:stability}). A cold extended gas reservoir would leave features imprinted on the cocoon spectrum from the large column of warm and cold neutral gas ($N_{\rm H}\sim10^{23}\,\mathrm{cm}^{-2}$): damped Ly-$\alpha$ absorption, low-ionisation metal lines such as Na\,\textsc{i}\,D and Ca\,\textsc{ii}\,K, dust extinction, and potentially molecular features. These latter features may be strong due to the very large column density, though much weaker than at solar metallicity, since metallicities are quite low \citep{Tripodi2024,Maiolino2025,Nikopoulos2026}. This raises the issue of what LRD-like systems would look like at high metallicity. However, whether low-redshift, high-metallicity LRD-like systems could be heavily dust-enshrouded AGNs such as compact obscured nuclei \citep[CONs,][]{Aalto2015} with two or three orders of magnitude greater dust-to-gas ratios is beyond the scope of this paper.

Short-term AGN variability is suppressed because the cocoon smooths intrinsic variations over the photon diffusion time, $t_{\rm var}\gtrsim 1\,\mathrm{yr}\,(\tau_{e^-}/10)(R/10^{17}\,\mathrm{cm})$, consistent with multi-epoch JWST findings that LRDs are photometrically steady as a population \citep[e.g.][]{Furtak2025,Zhang2025b}. We predict that any residual variability should appear on longer timescales and most strongly in the \emph{\textit{un-scattered}} cores of the broad lines. 

Our framework can also address many further properties through more detailed modelling: the exact nature of the X-ray and radio emission (likely suppressed by the large gas column), spectro-polarimetry signatures, the population of lower-column cocooned AGNs, the nature of LRDs at low redshifts, cold gas absorption and dust extinction, and the emission from hot and/or warm dust near the dust sublimation zone. 
These, together with broader population studies, are left to future work.




\begin{acknowledgements}
We would like to thank Kenta Hotokezaka, Alex J. Cameron and Sune Toft for inspiring discussions.



\textbf{Code and data availability:}
The \sirocco code \citep{Matthews2025} and documentation are provided\footnote{https://github.com/sirocco-rt/sirocco}. The comparison with observed LRD makes use of the public JWST\ data collected as part of several observational programs with the NIRSpec spectrograph \citep{Jakobsen2022} with the following PIDs; 1181,
4106, 
4233, 
3567, 
5224, 
5664, 9433, 1345, 1213, 1215, 1286, 1180, 1208, 1212, 2561, 4750, 5015, 6368, 9214, 9223. These observations have been uniformly reduced and published as part of the Dawn JWST\ Archive\footnote{https://dawn-cph.github.io/dja} (DJA), \citep{DeGraaff2024_RUBIES,Heintz2025,Pollock2025} or in the case of MoM-BH*-1, kindly provided by R.~Naidu et~al\footnote{https://zenodo.org/records/15059215}. DJA is an initiative of the Cosmic Dawn Center (DAWN), which is funded by the Danish National Research Foundation under grant DNRF140. This work is based in part on observations made with the NASA/ESA/CSA \textit{James Webb} Space Telescope. The data were obtained from the Mikulski Archive for Space Telescopes (MAST) at the Space Telescope Science Institute, which is operated by the Association of Universities for Research in Astronomy, Inc., under NASA contract NAS 5-03127 for JWST.
\textbf{Funding:}
AS and KEH acknowledge support from Independent Research Fund Denmark (DFF) under grant 6219-00033B and 5251-00009B, respectively. AS, DW, RD, KEH and SAS acknowledge funding by the European Union (ERC, HEAVYMETAL, 101071865). Views and opinions expressed are, however, those of the authors only and do not necessarily reflect those of the European Union or the European Research Council. Neither the European Union nor the granting authority can be held responsible for them. RDN is co-funded by Villum Foundation. JHM acknowledges funding from a Royal Society University Research Fellowship (URFR1221062). CK and SAS were supported by the UK's Science and Technology Facilities Council (STFC, respectively grant ST/V001000/1 and ST/X00094X/1). 
\end{acknowledgements}

\bibliographystyle{aa}
\bibliography{refs}

\begin{appendix}


\section{Details of the observed sample}\label{app:obs_sample}

The LRD sample is defined as follows. An object must (i) exhibit either a Balmer break or a change in spectral slope across the Balmer-break wavelength, (ii) show broad H$\alpha$ emission confirmed by grating spectroscopy, and (iii) be consistent with an unresolved source in NIRCam/F444W based on PSF-convolved single-Sérsic fitting. This selection is intended to capture the diversity of the LRD population, from `V-shaped' SEDs to systems with prominent Balmer breaks, while minimizing potential contaminants. We first selected candidates from the DAWN JWST Archive (DJA; \citealt{DeGraaff2024_RUBIES,Heintz2025}), where we selected NIRSpec/PRISM spectra objects that show either (i) a strong Balmer break (quantified as $B_{4200/3500} = F_{4100-4300\text{\AA}}/F_{3400-3600\text{\AA}} > 1.2$ following \citealt{Binggeli2019}), or (ii) a `V-shaped' continuum with a change in spectral slope ($\beta_{\rm UV}<-0.2$, $\beta_{\rm opt}>0$, $\beta_{\rm opt}-\beta_{\rm UV}>0.5$ at $2\sigma$ following \citealt{deGraaff2025b}). The UV and optical slopes are power-law fits to $F_\lambda$ over rest 1200–3645 Å and 3645–7000 Å in line-free regions with uncertainties on the slopes taken from these power-law fits. For objects without PRISM spectra, we apply the same slope criteria, with $\beta_{\rm UV}$ from photometry and $\beta_{\rm opt}$ from line-corrected photometry or the G395M continuum. 
When using the expression `Balmer break strength', we refer to this $B_{4200/3500}$, such that $B_{4200/3500}>0$ implies explicitly Balmer break objects. For inclusion in the final LRD sample, we further required a $\Delta \mathrm{BIC}>10$ (Bayesian information criterion) in favour of a broad-line component and consistent with an unresolved F444W morphology based on single Sersic-profile. 

\begin{figure*}
\begin{center}
    \includegraphics[angle=0, width=0.83\linewidth]{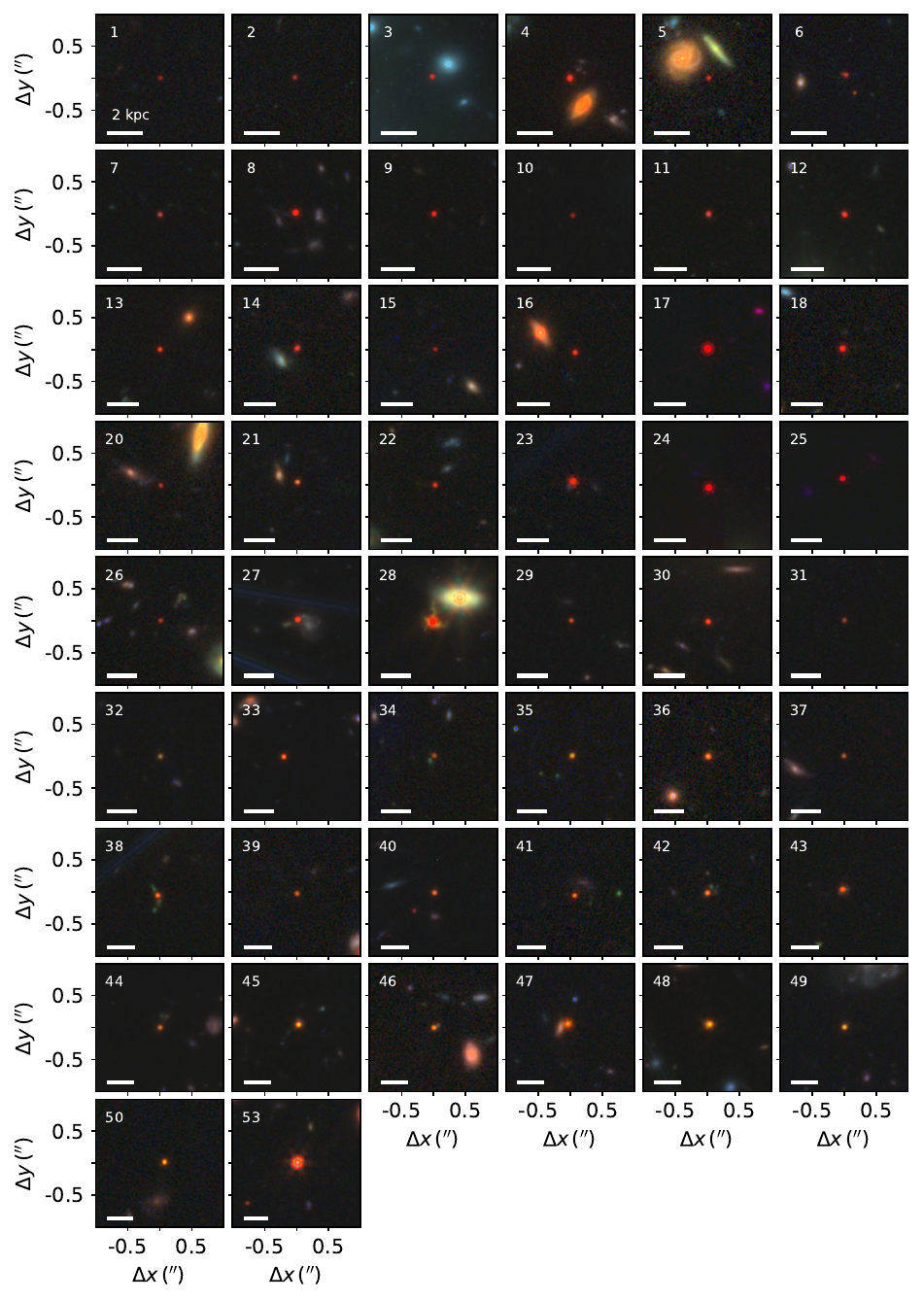}
\end{center}
\vspace{-0.6cm}
\caption{{JWST/NIRCam images of the observed sample}. All objects are compact and red, with occasional extended blue emission; three objects did not have full filter information needed for this image.}
\label{fig:images}
\end{figure*}


\begin{table}[!t]
\caption{NIRSpec target sample.}
\label{tab:msaid_targets}
 
\centering
\small
\renewcommand{\arraystretch}{1.1}
\setlength{\tabcolsep}{2pt}
 
\begin{tabular}{@{}r l l r r@{}}
\hline
ID & $z$ & MSAID & $\alpha$ (deg) & $\delta$ (deg) \\
\hline
1  & 7.1989 & rubies-uds31\_4233\_981721             &  34.3724 &  -5.1159 \\
2  & 7.0935 & mom-uds02\_5224\_920396                &  34.2478 &  -5.1520 \\
3  & 7.0370 & abell2744-spurs02\_9214\_41            &   3.5835 & -30.3967 \\
4  & 6.9831 & rubies-egs61\_4233\_55604              & 214.9830 &  52.9560 \\
5  & 6.7754 & rubies-uds42\_4233\_807469             &  34.3761 &  -5.3104 \\
6  & 6.7596 & jades-gdn2\_1181\_954                  & 189.1520 &  62.2596 \\
7  & 6.7279 & egs-nelsonx\_4106\_47962               & 214.8925 &  52.8569 \\
8  & 6.6849 & rubies-egs63\_4233\_49140              & 214.8922 &  52.8774 \\
9  & 6.3468 & uncover-61\_2561\_24968$^*$            &   3.6206 & -30.4000 \\
10 & 6.2230 & glimpse-obs01b\_9223\_5536             & 342.2563 & -44.5601 \\
11 & 5.8699 & jades-gdn\_1181\_38147                 & 189.2707 &  62.1484 \\
12 & 5.6840 & rubies-uds22\_4233\_139709             &  34.2960 &  -5.1499 \\
13 & 5.6814 & rubies-egs52\_4233\_37124              & 214.9910 &  52.9165 \\
14 & 5.6718 & rubies-uds3\_4233\_47509               &  34.2646 &  -5.2326 \\
15 & 5.6225 & ceers\_1345\_746                       & 214.8091 &  52.8685 \\
16 & 5.5819 & rubies-uds23\_4233\_172350             &  34.3690 &  -5.1039 \\
17 & 5.5350 & fresco-gn-5664\_9771$^*$               & 189.2810 &  62.2473 \\
18 & 5.5160 & capers-udsp1\_6368\_23419$^*$          &  34.4711 &  -5.1905 \\
19 & 5.4820 & fresco-gs-5664\_13971$^*$              &  53.1386 & -27.7903 \\
20 & 5.4421 & rubies-uds33\_4233\_29813              &  34.4534 &  -5.2707 \\
21 & 5.3580 & fresco-gn-5664\_16813$^*$              & 189.1793 &  62.2925 \\
22 & 5.2858 & rubies-egs51\_4233\_926125             & 215.1371 &  52.9886 \\
23 & 5.2771 & rubies-egs53\_4233\_42046              & 214.7954 &  52.7888 \\
24 & 5.2410 & fresco-gn-5664\_12839$^*$              & 189.3443 &  62.2634 \\
25 & 5.0850 & fresco-gn-5664\_15498$^*$              & 189.2855 &  62.2807 \\
26 & 5.0795 & rubies-egs61\_4233\_61496              & 214.9724 &  52.9622 \\
27 & 5.0769 & jades-gds06\_1286\_159717$^*$          &  53.0975 & -27.9013 \\
28 & 5.0389 & jades-gdn\_1181\_68797$^*$             & 189.2291 &  62.1462 \\
29 & 5.0068 & macs0416\_1208\_3104035                &  64.0562 & -24.1135 \\
30 & 4.9520 & valentino-egs\_3567\_42232             & 214.8868 &  52.8554 \\
31 & 4.8486 & jades-gdn\_1181\_39353$^*$             & 189.2940 &  62.1531 \\
32 & 4.8214 & jades-gds02\_1286\_38562$^*$           &  53.1359 & -27.8716 \\
33 & 4.8200 & jades-gds-w09\_1212\_2993              &  53.0888 & -27.7382 \\
34 & 4.8004 & egs-nelsonx\_4106\_71365               & 214.8955 &  52.9067 \\
35 & 4.7145 & rubies-uds21\_4233\_182791             &  34.2138 &  -5.0870 \\
36 & 4.6867 & gto-wide-uds14\_1215\_4446$^*$         &  34.2707 &  -5.2177 \\
37 & 4.5333 & capers-egs47\_6368\_19300              & 215.0221 &  52.9208 \\
38 & 4.2239 & rubies-egs63\_4233\_28812              & 214.9241 &  52.8491 \\
39 & 4.1478 & rubies-uds22\_4233\_119957             &  34.2689 &  -5.1767 \\
40 & 4.1327 & jades-gdn09\_1181\_73488$^*$           & 189.1974 &  62.1772 \\
41 & 4.1293 & rubies-uds1\_4233\_31747               &  34.2238 &  -5.2602 \\
42 & 4.1159 & rubies-uds23\_4233\_167741             &  34.3346 &  -5.1103 \\
43 & 4.0527 & gto-wide-uds13\_1215\_4273             &  34.3635 &  -5.1770 \\
44 & 3.9375 & jades-gds-wide\_1180\_13329            &  53.1390 & -27.7844 \\
45 & 3.7473 & glimpse-obs01b\_9223\_43084            & 342.2334 & -44.5391 \\
46 & 3.6851 & rubies-uds42\_4233\_23438              &  34.3383 & -5.2809 \\
47 & 3.5454 & rubies-uds31\_4233\_154183             &  34.4107 & -5.1297 \\
48 & 3.5026 & glimpse-obs02\_9223\_17775             & 342.2008 & -44.5437 \\
49 & 3.4300 & jades-gdn198\_1181\_53501$^*$          & 189.2951 &  62.1936 \\
50 & 3.3538 & rubies-uds23\_4233\_144195             &  34.3252 & -5.1437 \\
51 & 3.1073 & rubies-uds1\_4233\_40579               &  34.2442 & -5.2459 \\
52 & 2.3203 & gto-wide-egs1\_1213\_2974$^*$          & 215.1267 &  53.0484 \\
53 & 2.2603 & jades-gdn2\_1181\_28074                & 189.0646 &  62.2738 \\
\hline
\end{tabular}
\tablefoot{Redshifts, target identifiers, and sky coordinates for the 53 sources analysed here. Objects marked with an asterisk were observed with the G395H or G235H grating; all others were observed with G395M/G235M.}
\end{table}

The full comparison observational sample is summarized in Fig.~\ref{fig:images}, which shows JWST/NIRCam imaging and in Table~\ref{tab:msaid_targets}, which lists the coordinates, MSAID and spectroscopic setup. 
Table~\ref{tab:msaid_models_bic} further reports the best-fitting line-profile models (with the sole exception of rubies-uds1\_4233\_40579, where only the extreme wings of \ha are covered in the grating making any model distinction insignificant). For comparison, \citet{Scholtz2026} finds in their sample that 7 out of 7 X-ray detected AGN are better described with one or two broad Gaussian lines, whereas only 2 out of 25 LRD-like objects are better fit within the Gaussian-family parameterization. The only significant Gaussian-favoured object jades-gdn09\_1181\_73488 is notably an object where we achieve a best-fit with an exponential profile. We emphasise that exponential and Lorentzian line-profiles can be hard to discriminate at low SNR, and that providing significance against H$\alpha$ SNR is useful for cross-comparisons. When converting flux to luminosities we assume the flat $\Lambda$CDM-dominated concordance cosmological model \citep{2020_Planck}. 

We show an example fit, residual and posterior of the highest SNR \ha line Glimpse 17775 in Fig.~\ref{fig:Glimpse}. We note that these are examples of the predicted asymmetric exponential system, where significant improvements to the $\chi^2$ in the wings of the broad line follow from permitting a steeper profile on the blue side. \citet{Kokorev2025} previously noted asymmetry in the Glimpse 17775 \ha line as likely caused by blueshifted Balmer absorption, but we emphasise that even allowing for absorption in the symmetric wings provides a worse fit, as the asymmetry is a phenomenon extending to the extreme velocity-wings of the line at $\pm3000\mathrm{\,km\,s}^{-1}$. It is worth emphasising that this asymmetry is significant despite freedom in fitting the broad-line center offset from the host or narrow-line redshift. For most objects the large degeneracy in asymmetry and the broad-line centroid center inflates the error bars and permits the individual objects to be consistent with symmetry within 2$\sigma$ (although we note all these LRDs show steeper blueward slopes if we do not permit the centroid to drift redward by $\gtrsim 100\,\mathrm{km\,s^{-1}}$). We note that the fits are based on a sample with a substantial redshift range, which places \ha at different observed wavelengths. We also tested the robustness of extraction by comparing the results of different observations of the same objects that had multiple gratings, or IFU vs. slit or multiple visits. These generally sample different locations on the shutter or IFU. Our fits are consistent across these different datasets. 

\onecolumn
{\renewcommand{\arraystretch}{1.05}
\begin{longtable}{r l r r r r r r}
\caption{\label{tab:msaid_models_bic}Model comparison for the H$\alpha$ line profile.}\\
\hline\hline
ID & Best-fit model & $\exp$ & $\exp_{\rm abs}$ & gaus & gaus$_{\rm abs}$ & $\exp_{\rm asym}$ & $\exp_{\rm asym,abs}$ \\
\hline
\endfirsthead
\caption{continued.}\\
\hline\hline
ID & Best-fit model & $\exp$ & $\exp_{\rm abs}$ & gaus & gaus$_{\rm abs}$ & $\exp_{\rm asym}$ & $\exp_{\rm asym,abs}$ \\
\hline
\endhead
\hline
\endfoot
\hline
\multicolumn{8}{@{}p{12.3cm}@{}}{\small\textbf{Notes.} For each source, the table lists the preferred model and the $\Delta$BIC values of the alternative models relative to that preferred model. A value of --- indicates the best-fit model for that source. Rubies~4233-40579 is equally well fit by all models because it lies near the edge of the grating.}\\
\endlastfoot
1  & $\exp_{\rm abs}$      &   11.5 & ---  &     5.2 &    19.0 &    15.6 &   4.4 \\
2  & $\exp$                & ---    & 14.1 &    -0.2 &    11.5 &     5.2 &  13.6 \\
3  & $\exp$                & ---    & 17.3 &    -0.7 &    34.0 &     5.8 &  22.8 \\
4  & $\exp_{\rm abs}$      &  781.9 & ---  &   645.2 &    21.7 &  1048.7 &   2.7 \\
5  & $\exp$                & ---    & 12.5 &     7.3 &    34.7 &     5.6 &  18.8 \\
6  & $\exp$                & ---    & 12.2 &     4.7 &    16.3 &     9.1 &  16.6 \\
7  & $\exp_{\rm abs}$      &   13.1 & ---  &     6.4 &     0.3 &    17.1 &   4.6 \\
8  & $\exp_{\rm abs}$      &  233.2 & ---  &   243.4 &     5.9 &   239.2 &   6.1 \\
9  & $\mathrm{gaus}$       &    2.4 &  3.4 & ---     &     2.6 &     0.7 &   8.3 \\
10 & $\exp$                & ---    & 11.2 &     0.3 &    11.6 &     3.3 &  14.8 \\
11 & $\exp_{\rm abs}$      &    2.5 & ---  &     1.5 &    12.9 &     7.4 &   5.8 \\
12 & $\exp$                & ---    & 14.0 &     4.0 &    18.3 &     4.9 &  18.6 \\
13 & $\exp$                & ---    & 17.1 &     0.7 &    14.6 &     0.2 &  17.9 \\
14 & $\exp$                & ---    &  7.1 &     0.3 &    19.9 &     3.9 &  13.4 \\
15 & $\exp$                & ---    & 17.3 &     1.2 &    24.8 &    6.3 &  20.9 \\
16 & $\exp_{\rm abs}$      &   46.9 & ---  &    74.7 &    10.7 &    23.1 &   5.6 \\
17 & $\exp_{\rm asym,abs}$ & 7170.5 &  9.8 &  9533.6 &  2102.0 &  5380.9 & ---   \\
18 & $\mathrm{gaus}$       &    1.8 &  8.7 & ---     &    14.0 &     7.4 &  10.0 \\
19 & $\exp_{\rm abs}$      & 1038.7 & ---  &   948.9 &    71.3 &   902.7 &   6.6 \\
20 & $\exp$                & ---    & 11.5 &     1.5 &     4.6 &     3.2 &  11.0 \\
21 & $\exp_{\rm abs}$      &  102.3 & ---  &   608.4 &    80.5 &   109.0 &  35.7 \\
22 & $\exp$                & ---    &  1.6 &     9.3 &    13.3 &     0.2 &  12.4 \\
23 & $\exp_{\rm asym,abs}$ &  975.6 &  2.1 &   839.7 &    72.4 &   958.4 & ---   \\
24 & $\exp_{\rm abs}$      &   87.0 & ---  &   363.9 &   122.3 &    84.4 &  44.0 \\
25 & $\exp_{\rm abs}$      & 1130.8 & ---  &  1132.8 &    28.5 &   954.7 &   4.0 \\
26 & $\exp$                & ---    & 12.7 &    -0.3 &    14.6 &     5.9 &  17.8 \\
27 & $\exp_{\rm abs}$      &  255.3 & ---  &   253.9 &    18.1 &   261.9 &   6.8 \\
28 & $\exp_{\rm asym,abs}$ &  720.7 &  1.2 &   484.0 &    29.1 &   618.6 & ---   \\
29 & $\exp$                & ---    &  9.0 &     0.6 &    13.1 &     5.1 &  18.2 \\
30 & $\exp_{\rm asym,abs}$ &    1.5 &  2.8 &    22.0 &    19.2 &     5.3 & ---   \\
31 & $\exp$                & ---    & 15.8 &     0.9 &    23.0 &     6.5 &  32.7 \\
32 & $\exp$                & ---    & 21.5 &    17.8 &    33.5 &    98.5 &  15.8 \\
33 & $\exp_{\rm asym,abs}$ &   71.0 & 10.5 &    67.4 &    11.2 &    25.7 & ---   \\
34 & $\exp$                & ---    & 13.4 &    -0.3 &    12.8 &     3.8 &  17.6 \\
35 & $\exp_{\rm asym,abs}$ &  197.0 &  0.8 &   156.6 &    33.8 &   201.5 & ---   \\
36 & $\exp$                & ---    & 15.0 &     1.7 &    12.1 &     6.6 &  16.2 \\
37 & $\mathrm{gaus}$       &    0.8 & 14.6 & ---     &    14.7 &     8.4 &  21.5 \\
38 & $\exp_{\rm abs}$      &   96.9 & ---  &    39.6 &     5.1 &    56.3 &   5.4 \\
39 & $\exp$                & ---    & 11.7 &     1.6 &    14.8 &     5.3 &  20.5 \\
40 & $\exp_{\rm abs}$      &   59.3 & ---  &   131.7 &    28.5 &    65.6 &  32.8 \\
41 & $\exp_{\rm abs}$      &    2.1 & ---  &     1.2 &     0.5 &     5.0 &   4.5 \\
42 & $\exp_{\rm asym,abs}$ &  153.5 & 162.5 &  187.0 &   160.4 &   148.4 & ---   \\
43 & $\exp$                & ---    & 18.0 &    -0.1 &    18.0 &     5.8 &  24.6 \\
44 & $\exp_{\rm asym}$     &    3.3 &  3.3 &     5.5 &    13.7 & ---     &  10.2 \\
45 & $\exp_{\rm asym}$     &    3.4 & 16.8 &    30.2 &    25.1 & ---     &  14.4 \\
46 & $\exp$                & ---    & 14.5 &     0.4 &    15.5 &     8.6 &  19.8 \\
47 & $\exp_{\rm asym,abs}$ &    9.8 &  1.5 &    12.8 &    10.3 &    11.8 & ---   \\
48 & $\exp_{\rm asym,abs}$ & 705.2 & 30.2 & 723.7 & 133.0 & 482.0 & ---   \\
49 & $\exp$                & ---    &  5.7 &     9.6 &     6.5 &     5.9 &   4.0 \\
50 & $\mathrm{gaus}$       &    0.7 &  8.2 & ---     &     9.0 &     5.0 &  13.7 \\
51 & ---                   & ---    & ---  & ---     & ---     & ---     & ---   \\
52 & $\exp$                & ---    & 12.4 &    -1.0 &    11.9 &     6.0 &  20.4 \\
53 & $\exp_{\rm asym,abs}$ & 13170.7 & 152.5 & 11327.5 &   567.8 & 10210.5 & ---   \\
\hline
\end{longtable}
}

\twocolumn

\section{Sirocco model structure}\label{app:sirocco_modelling}\label{app:materials}

\begin{figure*}
\vspace{-0.2cm}
\begin{center}
    \includegraphics[angle=0,width=\textwidth]{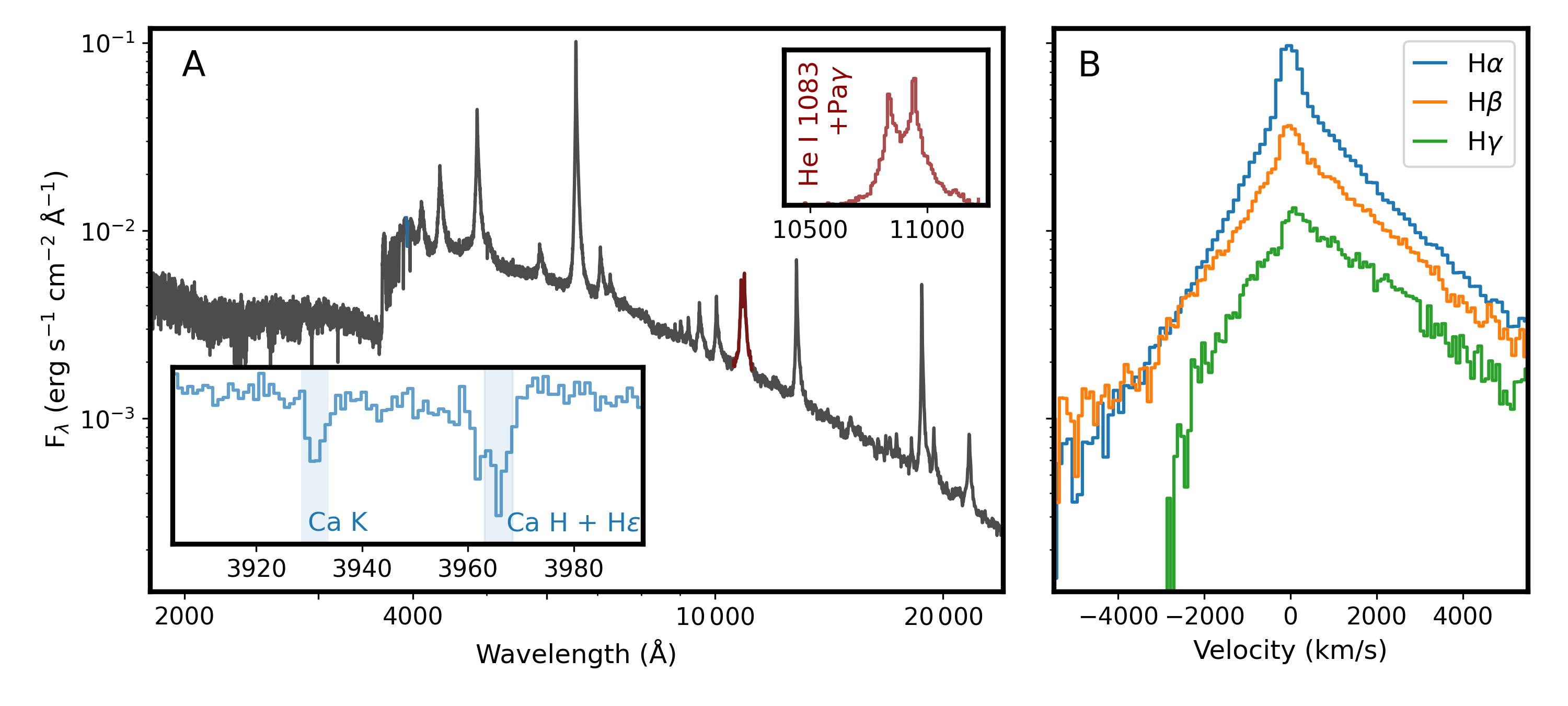}
\end{center}
\vspace{-0.9cm}
\caption{{Example model spectrum of an LRD-like object computed with \sirocco}. Prominent H and He emission lines as well as \caii H and K lines are highlighted. Even though the model is kinematically spherical, it reproduces several characteristic features of observed LRDs, including the Balmer break, broad H and He emission lines with exponential line profiles. 
Conditions sufficient to produce a Balmer break and self-absorption naturally produce electron-scattering--broadened lines. Despite electron scattering being nearly wavelength independent, the hydrogen line profiles are not identical because wavelength-dependent continuum opacity modifies the emergent wings. The spherical outflow also produces strongly asymmetric profiles. The parameters for this model are outlined in Table~\ref{tab:sirocco_params} and the model is viewed from a distance of 100\,pc.}
\label{fig:baseline}
\end{figure*}

\begin{figure}   
\begin{center}
\includegraphics[angle=0,width=0.8\linewidth,viewport=470 15 710 315,clip=]{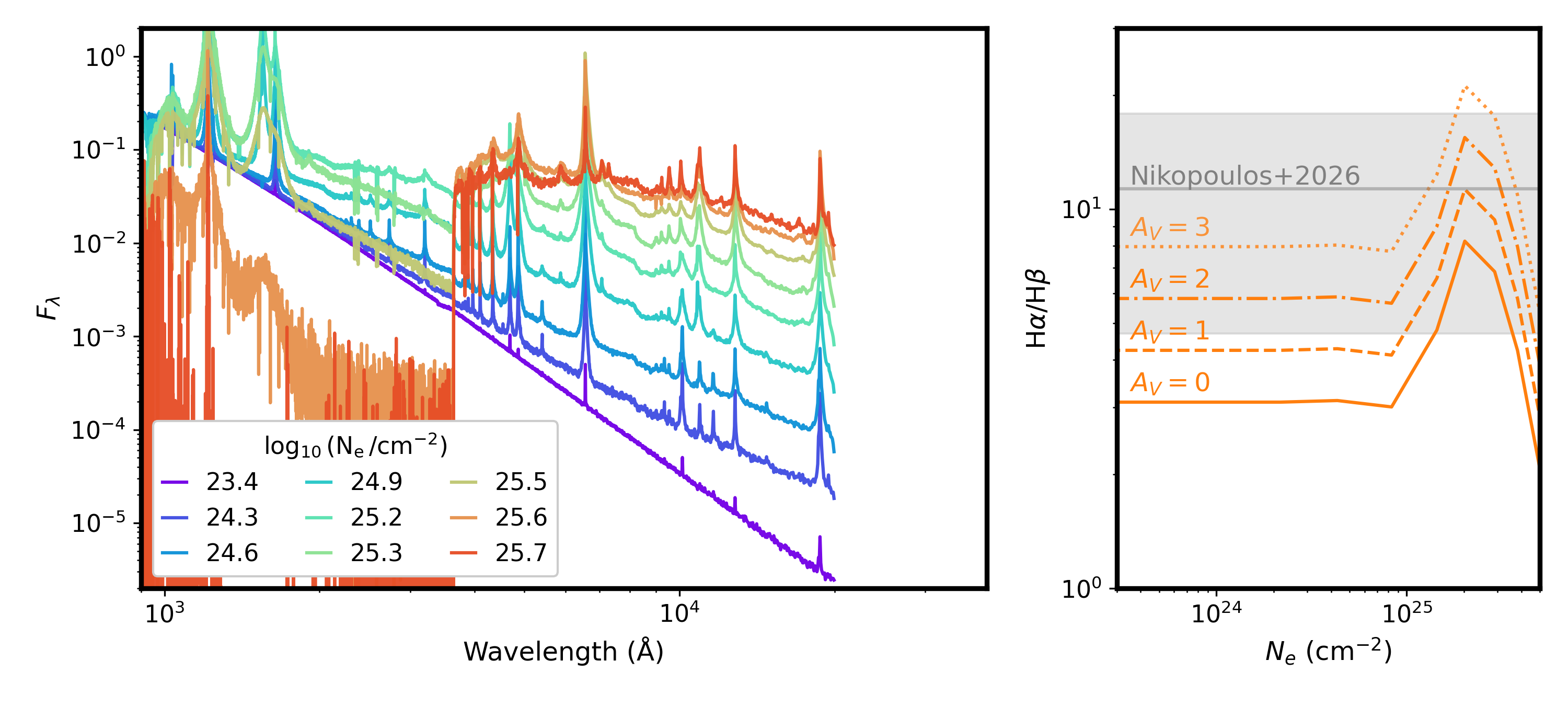}
\vspace{-0.2cm}
\caption{{Line ratios of $H\alpha/H\beta$ from a sequence of \sirocco spherical models with different column densities in a homologous wind}. Electron-scattering wings require a lower column density ($N_e\gtrsim 10^{24}\,\mathrm{cm^{-2}}$) compared to the column densities needed to produce Balmer breaks and Balmer line absorption ($N_H\gtrsim 10^{25}\,\mathrm{cm^{-2}}$). The high-column density regime with all three effects is characterised by strong deviations from Case~B line ratios, where the Balmer decrement could be as interpreted as $A_V\sim3$ if due to dust extinction. The mean and spread of the observed Balmer decrement from \citet{Nikopoulos2025} are indicated with the shaded region. }
\label{fig:density-sequence-caseB}
\end{center}
\end{figure}

\begin{figure*}
\vspace{-0.2cm}
\begin{center}
    \includegraphics[angle=0,width=1\textwidth]{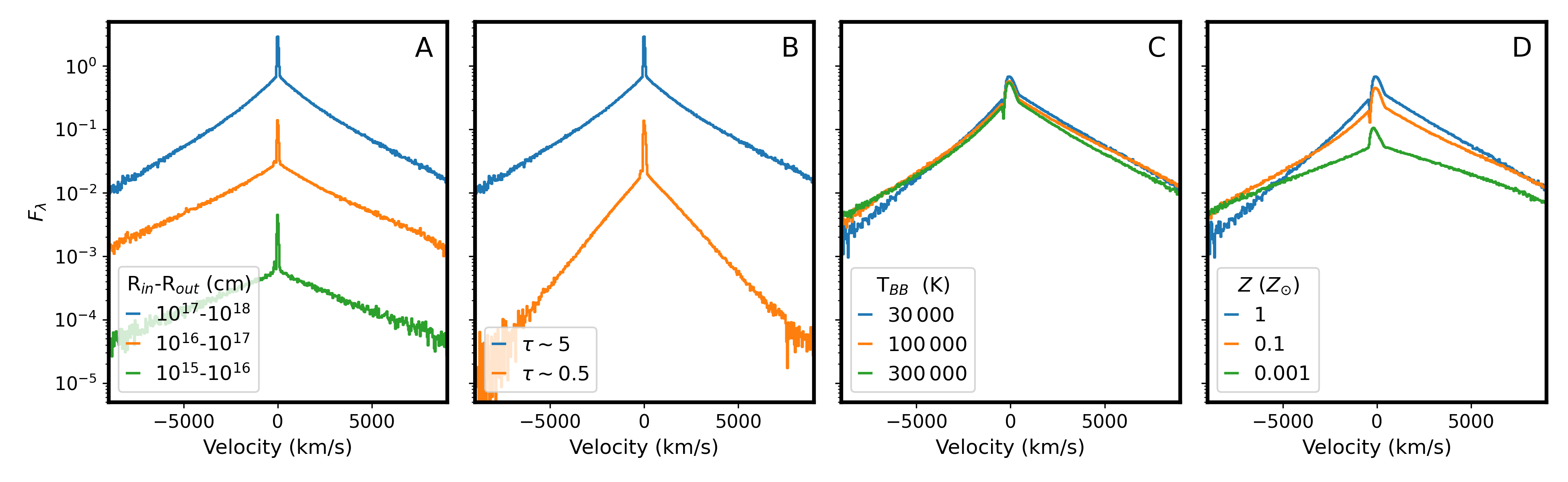}
\end{center}
\vspace{-0.7cm}
\caption{{Example \ha lines in \sirocco spherical models}. Each panel shows variations in \emph{A}) the characteristic radius (for a fixed hydrogen column density), \emph{B}) the density (or equivalently the optical depth), \emph{C}) temperature of the input radiation field and \emph{D}) metallicity of the cocoon. The two leftmost panels show order-of-magnitude differences in line luminosities because larger densities imply larger recombination-line fluxes. }
\label{fig:variations_of_parameters}
\end{figure*}

\begin{figure}
    \includegraphics[angle=0,width=\linewidth]{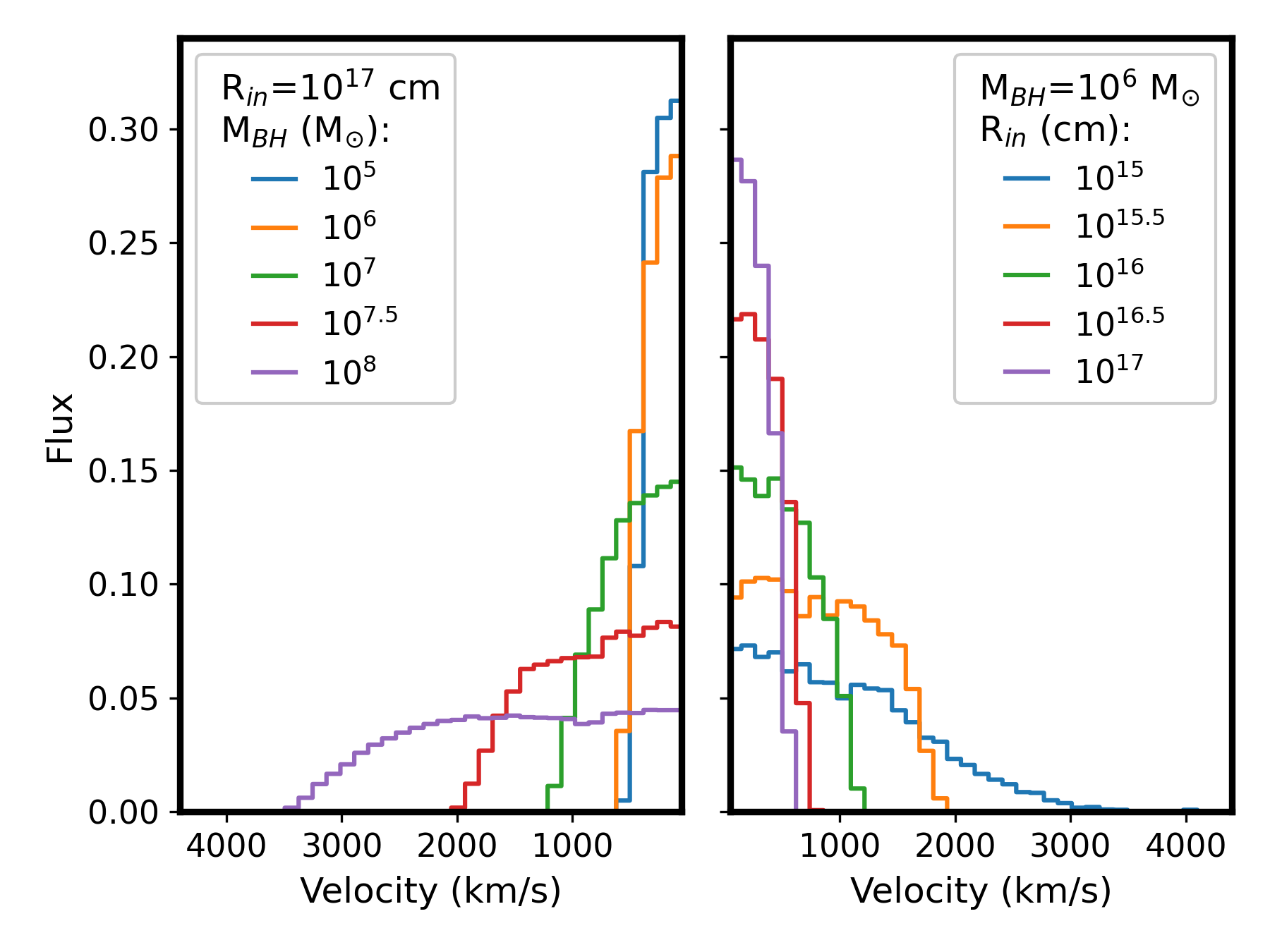}
\caption{{\ha profile without scattering as a function of SMBH mass (\emph{left}) and gas cocoon proximity (\emph{right}).} The central (un-scattered) \ha line becomes increasingly broad with increasing SMBH mass. A similar broadening effect is produced by shrinking $R_{in}$, because the relevant Keplerian velocity, \(v_\mathrm{Kep}\), is \(v_\mathrm{Kep}\propto M_{BH}^{1/2} R_{in}^{-1/2}\). }\label{fig:Kepler}
\end{figure}

\begin{figure}
    \includegraphics[angle=0,width=\linewidth]{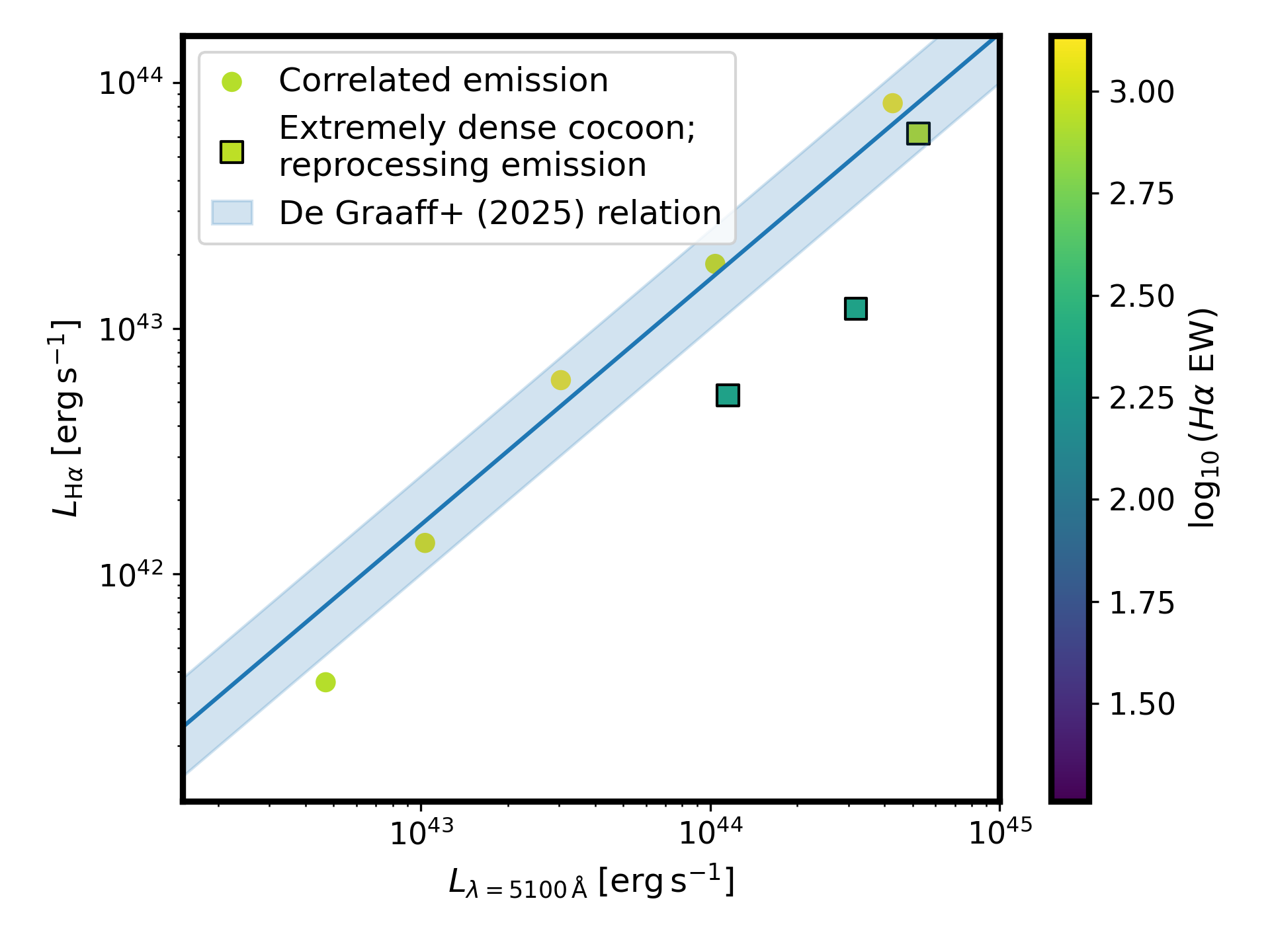}
\caption{{\ha line luminosity as a function of continuum luminosity at 5100\,\text{\AA}.} These \sirocco model outputs are correlated with similar offset and slope as the observational sample in \citet{deGraaff2025b}. The drop-off found by \citet{deGraaff2025b} is also reproduced for the extremely dense cocoons, when the recombination lines begin to be reprocessed. }\label{fig:Halpha_cont}
\end{figure}

\begin{figure*}
    \includegraphics[angle=0,width=\textwidth]{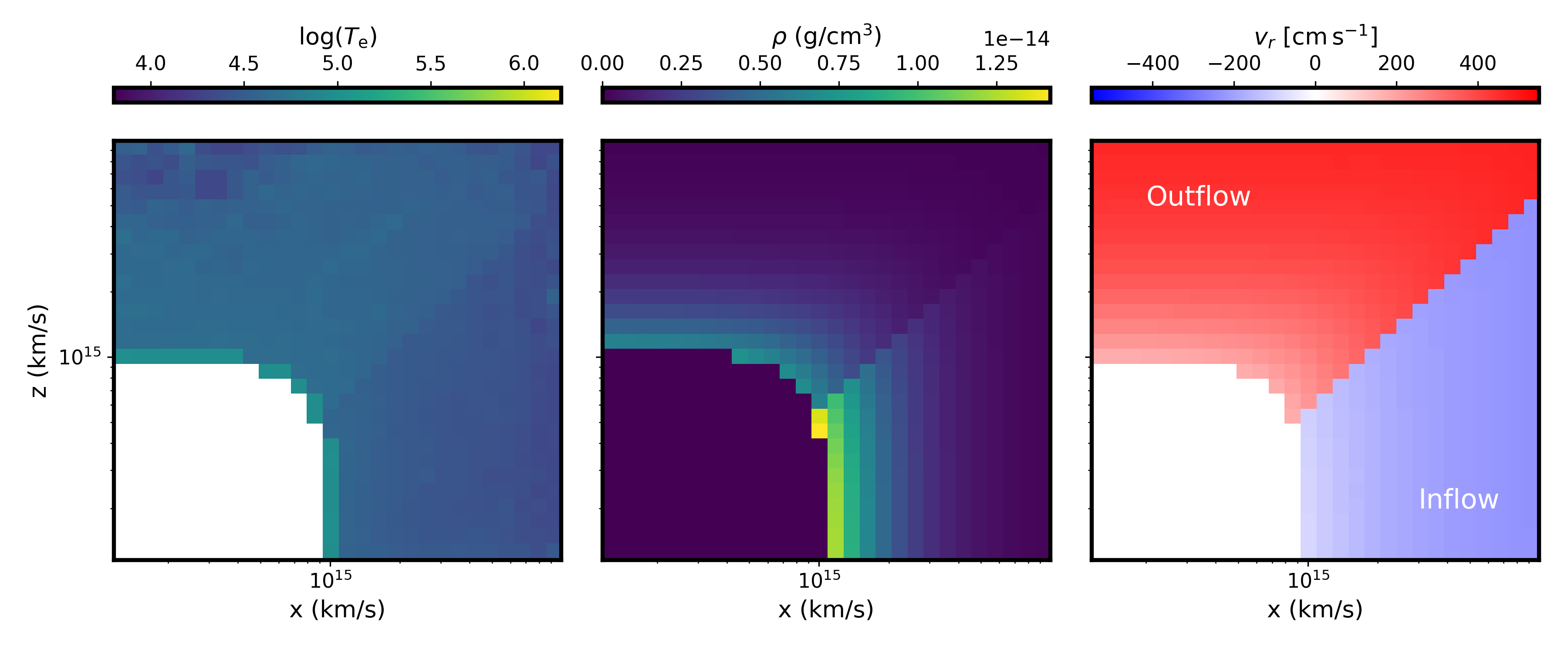}
\caption{{2d map of properties for an inflow/outflow model.} \emph{left}: Electron temperature map in the 2d plane. The model is based on the input density and velocity fields shown (\emph{centre} and \emph{right} respectively), where the inflowing equatorial gas is denser and slower than the outflowing polar gas.}\label{fig:wind_components}
\end{figure*}

Figure~\ref{fig:baseline} illustrates the ensemble of observables produced by the model, including the continuum, prominent broad permitted lines with variable line-shapes and high-equivalent-width absorption features (e.g.\ Ca\,\textsc{ii}) that can form in a cold cocoon, as observed in `the Egg' LRD \citep{Lin2025}. \sirocco simulations frequently produce a \hei~1083\,nm absorption superposed on broad \hei~1083\,nm emission, as shown in Fig.~\ref{fig:hei}, together with \feii features and absorption from \nai D doublet, while the atomic data for K is not included and thus lines of this element are not explored. Further, Fig.~\ref{fig:density-sequence-caseB} shows the model spectra also reproduce observed deviations from Case~B recombination ratios expected under dense gas conditions \citep{Nikopoulos2025,DEugenio2025b,Yan2025}. We next describe the required atomic data, radiative transfer mode and typical parameters for such high-fidelity modelling.

\textit{Atomic data and radiative transfer mode:} The electron-scattering module in \sirocco has been validated both in previous studies \citep{Parkinson2025} and via direct comparison with the independent electron-scattering code of \citet{Rusakov2025}. Resonance scattering is included using an escape-probability formalism, but it contributes only a minor fraction of the total scattering in all models considered. We use the `hybrid macro-atom' radiative transfer (RT) mode in \sirocco \citep{Matthews2015,Matthews2025}. In this mode, we adopt the full macro-atom formalism \citep{Lucy2002,Lucy2003,Sim2005,Matthews2015} for H and He, which includes all the important bound-free and bound-bound processes, but treat metals as `simple atoms', wherein recombination is less accurately modelled and line transfer is treated with a two-level approximation. We employ a thermal-trapping method for anisotropic scattering and note that the Sobolev approximation treatment only applies to line transfer. As the interior beneath the cocoon is not thermally solved, photons that reach the inner boundary are reflected. We include the following elements within \sirocco: H, He, C, N, O, Ne, Na, Mg, Al, Si, S, Ar, Ca and Fe. H and He levels are included up to principal quantum numbers $n=40$ and $n=10$, respectively (providing high-fidelity for Balmer physics) and metals are treated with two-level approximation for line transfer, with cross-sections/lines/collisions from the bundled datasets. At low electron temperatures, the thermal balance is a competition between primarily bound-free heating and line cooling. As \sirocco omits some heating/cooling processes at low temperatures, the resulting temperature structure can differ from a more complete treatment. For a more detailed explanation of parameters, code assumptions and atomic data we refer the reader to \citet{Matthews2025}.

\begin{table*}
\centering
\caption{Parameters explored in the \sirocco model suite.}
\label{tab:sirocco_params}
 
\begin{tabular}{llll}
\hline
Parameter & Meaning & Fig.~\ref{fig:baseline} & Range explored \\
\hline
$M_{\rm BH}$ & Black-hole mass & N/A & $10^{5}{-}10^{8}\,\mathrm{M}_{\odot}$ \\
$L_{\rm BH}$ & Accretion luminosity & $10^{44}\,\mathrm{erg\,s^{-1}}$ & $10^{43}{-}10^{46}\,\mathrm{erg\,s^{-1}}$ \\
$T_{\rm BB}$ & Input blackbody temperature & $10^5\,\mathrm{K}$ & $6\times10^{3}{-}10^{7}\,\mathrm{K}$ \\\hline
$R_{\rm in}$ & Inner cocoon radius & $10^{16}\,\mathrm{cm}$ & $10^{15}{-}10^{17}\,\mathrm{cm}$ \\
$R_{\rm out}$ & Outer cocoon radius & $10^{17}\,\mathrm{cm}$ & $10{-}100\,R_{\rm in}$ \\
$\rho$ & Density at $R_{\rm in}$ & $2\cdot10^{-14}\,\mathrm{g\,cm^{-3}}$ & $10^{-18}{-}10^{-12}\,\mathrm{g\,cm^{-3}}$ \\
$f_V$ & Filling factor & $1$ & $0.01-1$ \\
$\alpha$ & Density power-law index & 2.5 & $1{-}4$ \\
$Z$ & Metallicity & $1\,\mathrm{Z}_{\odot}$ & $10^{-3}{-}1\,\mathrm{Z}_{\odot}$ \\
$v_{\rm out}$ & Mean outflow velocity & $300\,\mathrm{km\,s^{-1}}$ & $50{-}1000\,\mathrm{km\,s^{-1}}$ \\\hline
$\frac{\mathrm{d}\Omega_{\rm inflow}}{\mathrm{d}\Omega_{\rm outflow}}$ & Solid-angle ratio & 0 & $0{-}5$ \\
$\frac{v_{\rm inflow}}{v_{\rm outflow}}$ & Velocity ratio & -- & $0{-}5$ \\
$\frac{\rho_{\rm inflow}}{\rho_{\rm outflow}}$ & Density ratio & -- & $0{-}5$ \\
$\theta$ & Viewing angle & -- & $0\degree$--$90\degree$ \\
\hline
\end{tabular}
\tablefoot{The table lists the input parameters varied in the radiative-transfer calculations, together with the values adopted for the illustrative model shown in Fig.~\ref{fig:baseline}. The parameters are grouped into three sets: input properties from the SMBH ($M_{\rm BH}$, $L_{\rm BH}$, $T_{\rm BB}$), cocoon-structure parameters ($R_{\rm in}$, $R_{\rm out}$, $\rho$, $\alpha$, $Z$, $v_{\rm out}$), and multiple-component parameters describing inflow-to-outflow ratios. The illustrative model uses a stellar wind from 100 km\,s$^{-1}$ to 500 km\,s$^{-1}$ with an approximate density power law. It was chosen to demonstrate LRD observables including H and He emission-line shapes, a Balmer break, and metal absorption features.}
\end{table*}

\textit{Sirocco parameter space:} The cocoon should produce a column density of electrons, $N_e\sim10^{24}\,\mathrm{cm^{-2}}$, as suggested by the observationally inferred Thomson optical depths, $\tau_{e^-}=\sigma_TN_e$ of order unity \citep{Rusakov2025}. Expressed as a spherical mass-loss rate, this corresponds to $\dot{M} \sim  1\,\frac{M_{\odot}}{\mathrm{yr}}\,\left(\frac{1}{f_{e}}\right) \left(\frac{v}{10^7 \mathrm{cm/s}}\right)\left(\frac{R_{\mathrm{in}}}{10^{16} \mathrm{cm}}\right)^2$ (where $f_e$ is the mean degree of ionisation). This cocoon will be responsible both for reprocessing the incident radiation field into recombination-line emission and the broadening of these lines via electron scattering. 
The full list of parameters explored is summarised in Table~\ref{tab:sirocco_params}. We discuss the observable imprints of each parameter below, and then outline the preferred model. 

\textit{Surrounding gas cocoon}: We initialize a cocoon, spherical in terms of density, with an inner radius $R_{\mathrm{in}}\in[10^{15}$, $10^{17}]\,\mathrm{cm}$ and an outer radius much larger than this (i.e.\ $R_{\mathrm{out}}=10\,R_{\mathrm{in}}$ with larger distances providing similar results due to the steep density slopes). We explore a broad range of power-law slopes $1<\alpha<4$, but for our fiducial setup $\rho\propto r^{-2}$. 
Although mass-loss rate, velocity, and acceleration-structure are physically interdependent, we vary $\rho(r)$ and $v(r)$ independently to explore the observable consequences of each separately. The gas composition is a variable parameter: for our fiducial setup we adopt solar abundances, but also explore sub-solar metallicities, $Z=10^{-3}{-}1\,\mathrm{Z}_\odot$, since metal-poor gas cools less efficiently \citep{Sutherland1993} and thus yields broader exponential electron-scattering wings at fixed Thomson optical depth due to the higher electron temperatures. See Fig.~\ref{fig:variations_of_parameters} for the functional impact of these parameters on the line profile. We note that \sirocco can parametrically model unresolved clumping within cells, but we generally assume smooth winds, $f_{V}=\langle\rho\rangle/\rho_{cl}=1$. Clumping enhances density-squared processes, including recombination and collisional cooling, without a corresponding enhancement of the electron-scattering optical depth, which is set by the mean density. Such spatial inhomogeneity may also be physically motivated, for example by the observed escaping broad Ly-$\alpha$ emission \citep{Tang2026,Ji2026}.

\textit{Cocoon kinematics}: We parameterize the cocoon gas outflow with a velocity profile $v_{\mathrm{outflow}}(r)$, adopting a constant velocity, a $\beta$-law wind, or homologous expansion. The observations are relatively insensitive to the velocity parameterization, as both the exponential wings and Balmer line absorption are predominantly dictated by the average velocity. 
Beyond basic spherically symmetric velocity models, we look at multi-component axisymmetric models with independent velocities, densities, and solid-angles (Appendix~\ref{app:supp_text}). We also include a non-zero rotational component $v_\phi$, where we assume Keplerian motion around the SMBH, $v_{\phi}=\sqrt{GM_{BH}/r} \sin(\theta)$, where $\sin(\theta)$ accounts for the component of gravitational acceleration towards the rotational axis. For a sufficiently large SMBH or a nearby cocoon, this means that the dominant broadening of the un-scattered component is the Doppler motion (see Fig.~\ref{fig:Kepler}). 
We assume black hole masses within $M_{\mathrm{BH}}\sim10^{5-8}\,M_{\odot}$, consistent with revised estimates of the intrinsic \ha broad-line width after accounting for broadening from electron scattering \citep{Rusakov2025}. $M_{\mathrm{BH}}$ only sets the width of the Gaussian un-scattered broad-line core through Keplerian motion. It has no downstream dynamical or gravitational effects on the pre-imposed velocity-field. 

\textit{Input radiation field}: The total luminosity ($L \simeq 10^{43}{-}10^{45} \mathrm{\,erg\,s}^{-1}$), especially blueward of the Lyman limit, impacts the mass of ionised gas and the radius to which the cocoon is ionised and thus what the electron column density is. The ionised region extends far beyond the Strömgren radius, because a sustained population of hydrogen in $n=2$, maintained in photoionisation equilibrium with \ion{H}{ii}, allows Balmer-continuum photons to keep the gas ionised. The accretion emission is assumed to be a blackbody with characteristic temperature $T_{\mathrm{BB}} \sim 10^{5}\,\mathrm{K}$, isotropically emitted in $d\Omega=4\pi$. However, the spectral continuum are robust to order-of-magnitude variations in $T_{\mathrm{BB}}$, as the ionisation balance and radiation field are solved self-consistently within the gas, where the flow typically cools to O($10^4\,\mathrm{K}$).

\begin{figure*}
\begin{center}
    \includegraphics[angle=0,width=0.95\textwidth]{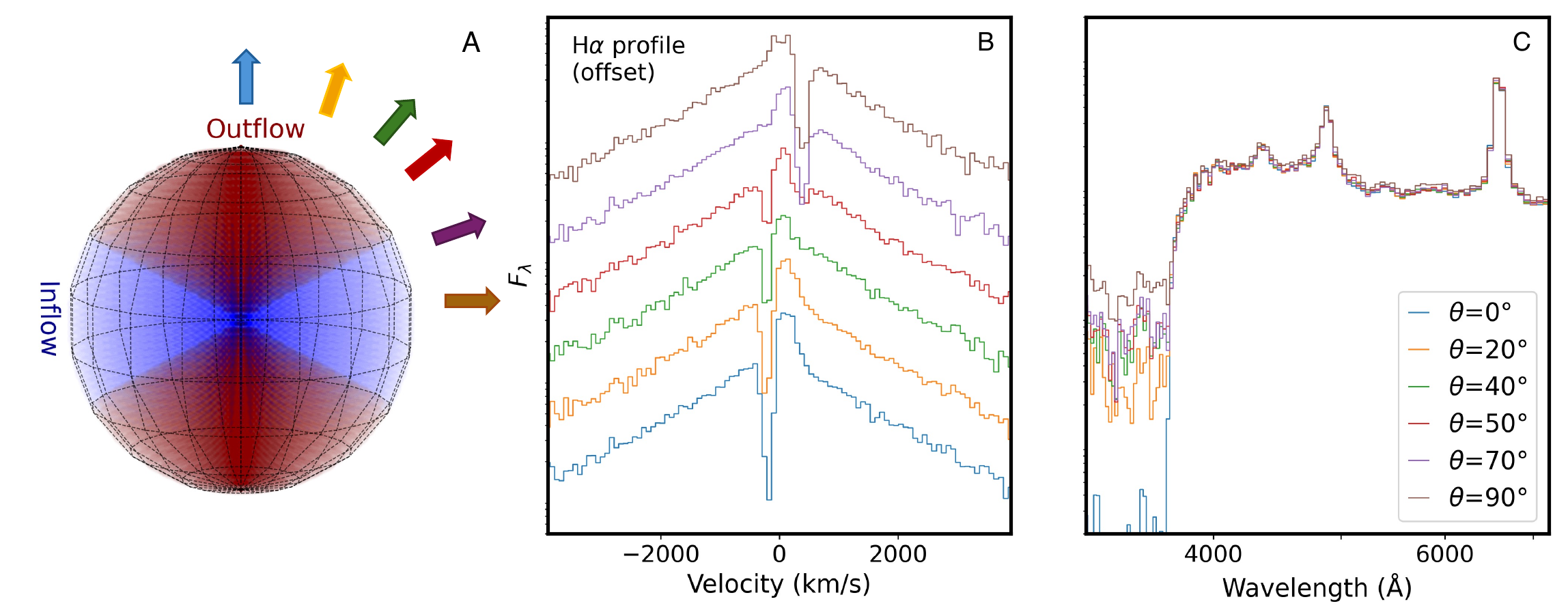}
\end{center}
\vspace{-0.7cm}
\caption{{Directional spectra from various viewing angles}. Velocity distribution of a cocoon with polar outflows and equatorial inflows with equal solid angle coverage (\emph{A}). The arrows indicate various observer directions from the polar axis. The corresponding \sirocco \ha line profile (\emph{B}) and continuum (\emph{C}) produced at these viewing angles. The central \ha profile is line-of-sight dependent with blue- and/or redshifted absorption, whereas the scattering-broadened line is highly symmetric from all viewing angles. The Balmer break strength is similarly line-of-sight dependent, whereas the red continuum of recombination emission is broadly similar across viewing angles (see also \citealt{Mascia2026}).}
\label{fig:viewing}
\end{figure*}

\textit{Multiple components}: As we explore broad ranges of geometries and cocoon structures, we use imported models in \sirocco. This accommodates multi-component set-ups with both inflows and outflows. The imported polar coordinate grids are uniformly sampled in $\theta$ and logarithmically in $r$, while the velocity-field reproduces the standard \sirocco 
coasting wind or homologous expansion. For the kinematic and geometric discussion in this paper, we are mainly concerned with the characteristic velocities in the scattering cocoon, which makes higher-order effects such as the parameterization of the radial velocity less important. In Fig.~\ref{fig:wind_components}, we show an example two-component model, where the left panel shows the output electron temperature at various spatial positions and highlights that although here we invoke a steep gradient between inflows and outflows, the resulting temperature solution does not display unphysical structures surrounding the shear layer. The central and right panels indicate the input density and velocity structure of the cocoon. 
We denote the viewing angle  $\theta$, ranging from a polar (0\degree) to equatorial (90\degree). For viewing angle effects, see Fig.~\ref{fig:UV} for continua and Fig.~\ref{fig:viewing} for continua and line shapes. While we deal here with a single equatorial/axial inflow/outflow model, many inflow/outflow components could be considered, such as when dealing with the multiple radial zones of convective bubbles (Appendix~\ref{app:supp_text}). Further, beyond the overall velocity flow one can also introduce more local perturbations to the imported velocity field: $\vec{v}=\vec{v}_{\rm flow}+\vec{\delta v}$. We primarily adopt the large-scale velocity field, $\vec{v}_{\rm flow}$, which is constrained by observations. To test the possible role of unresolved motions, we also consider stochastic or turbulent velocity components, $\vec{\delta v}$, and find that small (turbulent) perturbations are consistent with the observations. Their main effects are to increase absorption equivalent widths and modestly soften the continuum rollover beyond the Balmer break, suggesting that unresolved motions may affect detailed spectral modelling but are not unambiguously required by the current data.

\textit{Dust extinction and attenuation}: While the ionised cocoon providing electron scattering is inconsistent with significant dust (see Appendix~\ref{app:dust}), the diversity of red spectral slopes \citep[e.g.,][]{Barro2025}, and the conditions permitting dust at larger radii motivate the additional freedom to add reddening. In this work, we apply modest dust extinction ($A_V\lesssim1$) with Small Magellanic Cloud (SMC) extinction curve \citep{Gordon2003} as a standard high-redshift, low-metallicity dust prescription in Fig.\ref{fig:comparison_LRDs_Sirocco}, but otherwise show \sirocco model outputs without this added degree of freedom. If the dust is mixed with, or surrounds, the unresolved cocoon, the more appropriate description is an effective attenuation curve rather than a pure foreground-screen extinction curve, since scattering and geometry can affect the observed continuum. We therefore also tested a Calzetti-like attenuation law \citep{Calzetti2000}. Because this law is flatter than the SMC curve, it requires a somewhat larger $A_V$ to produce the same UV reddening, but for the modest dust columns considered here the resulting changes to the rest-optical \sirocco continuum slope are inconsequential for our analysis, especially given the lack of a tight prior on $A_V$. Similarly, adopting a flatter Milky-Way-like extinction curve would mainly affect the inferred UV attenuation and infrared energy budget, without materially changing the rest-optical continuum shapes considered here. If the observed LRD UV continuum is interpreted as a dust-reddened AGN continuum, the limited infrared excess of LRDs together with energy-conservation arguments, implies $A_V\lesssim 1{-}1.5$\citep{Chen2025}. This constraint could be relaxed somewhat if part of the observed UV emission has a distinct stellar origin, since that component need not be absorbed by the same LRD dust column. Further, recent modest rest-frame LRD NIR-excess detections have been interpreted as dust-emission with $A_V\sim0.8\pm0.2$ \citep{PerezGonzalez2026}. Ultimately, while modest reddening provides some additional flexibility in matching the spectral slopes, the characteristic gradual turnover and inflection are still produced by the underlying \sirocco model spectra. 

\textit{Preferred models}:  The diversity of our LRDs and LRD-like objects can be modelled with accreting black holes with luminosities ($10^{43}{-}10^{45}\,\mathrm{erg\,s^{-1}}$) and a surrounding cocoon of densities \(10^{6}{-}10^{10}\,\mathrm{cm}^{-3}\), initialized between $R=10^{15}{-}10^{17}$\,cm with a falling radial density profile, fiducially $\rho \propto r^{-2}$. The luminosity of the reprocessed red optical emission is set by the total ionised mass, which is noteworthy as it allows us to functionally decouple the question of LRD luminosity and spectral shape. In particular, a coherent re-scaling of i) cocoon size, ii) wind mass-loss rate ($\propto\rho/r^2$) and iii) input luminosity provides similar columns and thus spectral shapes, but a correspondingly rescaled output luminosity. Conversely, the Balmer break and scattering wings constrain the column of $n=2$ and ionised gas respectively. These are naturally connected, which is why picking the closest line-shape \sirocco models also provided a good match for the overall spectral continuum shape (i.e.\ as done in Fig.~\ref{fig:comparison_LRDs_Sirocco}). In this work, for comparisons to observed LRDs, we used i) the observed \ha line shapes to find the closest LRD model spectra in terms of kinematics and column of cocoon material and ii) rescaled the cocoon scale to match the observed luminosity. In future work, higher-detail fitting can be employed with a more complete LRD-template basis, which could assess the exact spectral shape of the continuum, the absorption of metals and several other parameters of interest.

\textit{Gas mass and temperature -- width of scattered broad lines:} 
The velocity width of the exponentially broadened lines is set primarily by electron scattering. The relevant parameters are the electron optical depth, $\tau_{e^-} = \sigma_T N_e$, which controls the number of scatterings, and the electron temperature, through $v_{th}=(2k_BT/m_e)^{1/2}$, which controls the frequency shift per scattering. The observed wing width therefore constrains a combination of $N_e$ and $T_e$. Continuum absorption provides a separate control on the emergent profile, because it preferentially removes multiply scattered photons and can suppress the broadest wings. This is why increasing column initially broadens the lines, while at very high columns bound-free absorption can narrow the emergent profile.

Varying the electron column density produces a range of different spectra; see Fig.~\ref{fig:density-sequence}. The column density required for an electron-scattering optical depth of order unity is comparable to or smaller than the hydrogen column densities implied by the Balmer break and Balmer self-absorption \citep{Naidu2025,Taylor2025} even for a modest ionisation fraction $f_e\gtrsim0.1$. Figure~\ref{fig:variations_of_parameters} (first panel) shows the exponentially broadened lines for a cocoon at three spatial scales for fixed $\tau_{e^-}\sim5$, which illustrates the cocoon's spatial scale is not tied to spectral shape but to the luminosity. The subtle differences in exponential slopes arise because higher-density models have greater absorption opacity and thus absorb photons in the scattering wings. Varying the cocoon's volume density (Fig.~\ref{fig:variations_of_parameters}, second panel) changes the optical depth and therefore the width of the exponential wings \citep{Huang2018}. High column densities can be achieved with relatively little mass. Given, i) $N_e=\int n_e(r)\, dr$, ii) total number of electrons $N_{e,tot} = \int n_e(r)\,dV$, and iii) total mass $M_{\rm cocoon}=N_{\rm tot}\mu m_{\rm H} / f_e$, we obtain 
\begin{equation}
    M_{\rm cocoon} \approx 1\,M_{\odot} \left( \frac{R_{in}}{10^{16}\,\mathrm{cm^{-2}}} \right)^2 \left( \frac{f_e}{1} \right)^{-1}\left( \frac{1-\alpha}{3-\alpha} \right) 
,\end{equation}
where the prefactor $\left( \frac{1-\alpha}{3-\alpha} \right)$ comes from integrating $n_e\propto r^{-\alpha}$ assuming $\alpha>3$. Conversely, for the wind-like structure, $\alpha=2$, we would get an additional factor of $R_{out}/R_{in}$. In all cases, however, only a relatively modest ionised cocoon mass is required compared to the SMBH mass to produce the observed characteristic features. This does not mean, however, that the full accreting gas cloud cannot be much more massive, and indeed, it must be to continue to feed the accretion onto the SMBH for longer than a few thousand years.

The velocity shift per scattering is harder to modify. Efficient cooling keeps the gas near $\lesssim10^4\,$K. For instance, varying the input radiation field from the SMBH under the blackbody ansatz (Fig.~\ref{fig:variations_of_parameters}, third panel) ultimately has only a minor effect on the resulting effective width of the lines. The equilibrium $T_e$ can be affected by gas metallicity, as cooling can become inefficient at $\gtrsim 10^5$ K in very low metallicity systems (Fig.~\ref{fig:variations_of_parameters}, fourth panel). It is possible in principle to constrain the electron-scattering optical depth independent of the line width based on the ratio of un-scattered to total line photons. If we trust this measure, we would infer the observed sample to have $\tau_{e^-}\sim2$. This optical depth then requires a relatively low temperature, $T_e\sim5{,}000$\,K, to match the observed widths. This relatively cool gas is consistent with independent temperature estimates from Fe\,\textsc{ii}, $T_{\mathrm{eff,\,Fe\,II}}\sim7{,}000$\,K \citep{Torralba2025} and Ca triplet absorption $T_{\mathrm{eff,\,Ca\,T}}\sim5{,}000$\,K \citep{Lin2025}.

\textit{SMBH mass -- Width of the un-scattered line:}
The un-scattered component of the broad lines likely still encodes information about the intrinsic Doppler motion of the gas and thus, by extension under the assumption that the dominant velocity field is virial/Keplerian, the SMBH mass (Fig.~\ref{fig:Kepler}). Radial outflow or inflow velocities, turbulence, and radiative-transfer effects can also contribute to the width of the un-scattered component. Massive-star winds provide an important comparison: massive star winds and Wolf--Rayet winds can reach terminal velocities of \(v_\infty\sim1000{-}3000\,{\rm km\,s^{-1}}\) \citep{Crowther2007,Puls2008}. Eruptive massive-star outflows are typically slower, of order a few hundred \({\rm km\,s^{-1}}\), although small amounts of material in some events may reach substantially higher velocities \citep{Smith2014,Davidson2020}. Line widths of several thousand \({\rm km\,s^{-1}}\) are not, by themselves, unique evidence for Keplerian motion around a SMBH. 
Nevertheless, the fitted un-scattered Gaussian-core component has a mean FWHM of \(\sim570\pm60\,{\rm km\,s^{-1}}\), and \(12\%\) of the sample has Gaussian-core FWHM \(>1000\,{\rm km\,s^{-1}}\). The mean width is comparable to the line-of-sight absorption velocities observed in LRDs, which are typically a few hundred \({\rm km\,s^{-1}}\) or less, but the high-width tail is difficult to attribute solely to the observed bulk radial outflows. Further, on the typical spatial scales of the associated ionised gas ($\sim10^{16}\,{\rm cm}$) a substantial Keplerian velocity would be expected for SMBH masses \(\gtrsim 10^6\,{\rm M_{\odot}}\). The Keplerian motion would only be expected to be negligible for relatively weak central potentials, i.e.\ with masses $\lesssim10^{5}\,{\rm M_{\odot}}$. Such low masses in comparison to the substantial observed luminosities of $10^{43}{-}10^{45}\,{\rm erg\,s^{-1}}$ would require highly super-Eddington accretion for an SMBH interpretation. We therefore interpret the un-scattered line width as being most naturally explained by gas moving in the SMBH potential, while noting that wind, inflow, and turbulent components may contribute additional width and are systematic uncertainties in the mass estimate. In general, the preference for a SMBH associated cocoon follows from the luminosity, compact emitting scale and fairly hot required central source, including strong, broadened \hei emission with superposed absorption.

\textbf{Sobolev criterion:} An operational assumption within \sirocco is the Sobolev approximation \citep{Sobolev1957,Sobolev1960}, which applies when the resonance region of the bound-bound interactions (the Sobolev length scale) is small compared to the other relevant scales of the system. In particular, the Sobolev length scale is set by the thermal speed of the ions, $v_{\rm ion,th}$, relative to the characteristic velocity gradient $|dv/dr|$. For a typical temperature ($10^4 \,\mathrm{K}$), a cocoon scale, $R$, and a homologous velocity scale with $v_{\rm outflow}=v(R)$, the Sobolev length in spherical symmetry becomes 
\begin{equation}
    l_s=\frac{v_{\rm ion,th}}{|dv/dr|}=10^{14} \,\mathrm{cm} \left(\frac{R}{10^{16}\,\mathrm{cm}}\right) \left(\frac{v_{\rm outflow}}{1000\,\mathrm{ km\,s}^{-1}}\right)^{-1} \left(\frac{v_{\rm ion,th}}{10\,\mathrm{ km\,s}^{-1}}\right)
\end{equation}
using $|dv/dr|\approx v_{\rm outflow}/R$ for a homologous outflow. In the more general case, including a coasting flow where $dv/dr = 0$, the term that sets how quickly one passes over the resonance region, e.g.\ $(dl/d\nu)_{r_s}=c/\nu_{lu} ((1-\mu^2)v/r+\mu^2dv/dr)_{r_s}^{-1}$, the velocity gradient is of the same order of magnitude, except for directions very close to radial \citep{Noebauer2019}. Importantly, the Sobolev scale remains much less than the characteristic scale of the system, the scale over which the ionisation or the temperature change, and it is smaller than the grid scale when we fiducially parameterize the cocoon structure at 10--50 radial positions. For small characteristic velocities, $\ll100\,\mathrm{km\,s}^{-1}$, we leave the microscopic regime for which \sirocco is designed. However, even in this regime the electron scattering (e.g.\ discussions of asymmetry, kinematics and geometry) would still be robust. 

Another, relevant comparison is the Sobolev length scale relative to the mean-free path due to electron-scattering, $l_{es}=(\sigma_T n_e)^{-1}$. In particular, as electron-scattering moves lines out of resonance, this may limit the \sirocco implementation of escape probabilities for Ly-$\alpha$ and by extension extended ionised zones sustained by photoionisation from the $n=2$ level. For a homologous flow the comparison reduces to a simple form,
\begin{equation}
    \frac{l_s}{l_{es}} = \sigma_T n_e \frac{v_{\rm ion,th}}{|dv/dr|}
    \simeq \tau_{es}\,\frac{v_{\rm ion,th}}{v_{\rm outflow}},
\end{equation}
so the Sobolev picture is internally consistent provided $\tau_{es}\lesssim v_{\rm outflow}/v_{\rm ion,th}$. At the electron-scattering photosphere and in regions associated with Balmer absorption this is satisfied with $l_s/l_{es}\lesssim0.04-0.26$ at $\tau_{es}=1$ across both the spherical and the 2D inflow--outflow model in Figs.~\ref{fig:cartoon_D} and \ref{fig:density-sequence}. The condition is stressed in the dense inner regions of the most optically thick models. Further work should address the effect of this limitation of the Sobolev approximation on the exact location and extent of partially ionised and neutral zones.

\section{Additional model diagnostics}\label{app:supp_text}

\subsection{Bound-free opacity in \sirocco}

The \sirocco models predict variations in broad-line widths between \ha and other hydrogen lines. In the main text, we mention that this is because the continuum bound--free absorption opacity varies with wavelength and the lower bound-free--opacity at \hb compared to \ha means less reprocessing and thus broader exponential lines (see Fig.~\ref{fig:comparison_LRDs_Sirocco}F). We reached this conclusion as follows. The linewidths may differ due to two effects: i) line formation at different cocoon depths, so that the lines see different column densities \citep{Chang2025}, or ii) continuum absorption is wavelength-dependent, affecting the opacity of the multiply scattered photons that populate the exponential wings differently for different lines \citep{Huang2018}. The strong wavelength-dependence of bound-free opacity ($\propto \lambda^{3}$ at energies above the photoionisation edge) means that lines closest to the relevant photoionisation edges (e.g.\ \ha is closer to the Paschen edge than \hb), suffer the largest continuum absorption losses to their wings. In the \sirocco-model spectra, these two causes of differences in linewidths can be separated explicitly, and we find that the continuum opacity dominates. The median \ha photon escapes farther out than \hb, but similar to or farther in than \pab (Fig.~\ref{fig:cartoon_D}, upper right). So while \ha photons traverse a similar or larger electron column than \pab, the \ha line is observed to be narrower than \pab in the sample objects and in the models due to the bound-free continuum opacity.

\subsection{Viewing angle in \sirocco model spectra}
A key parameter in \sirocco is the observer viewing angle. The red optical continuum arises from reprocessed radiation in the ionised gas cocoon, while the broad emission lines are shaped by electron scattering throughout this region. As a result, both the continuum and broad lines depend only weakly on viewing angle. This weak dependence implies that, at the resolution of NIRSpec/PRISM, the spectra can often be interpreted using simplified one-dimensional models (e.g.\ Fig.~\ref{fig:baseline}). In contrast, absorption features are strong diagnostics of the line of sight and therefore of the viewing angle. Higher-resolution grating spectra can thus constrain the geometry and kinematics of the flow. As shown in Fig.~\ref{fig:baseline}, this viewing-angle dependence explains the shifted absorption features in permitted lines. Balmer continuum absorption is likewise sensitive to the column of hydrogen atoms in the $n=2$ state, which varies significantly with viewing angle. This suggests a correlation between the centroid velocity of Balmer absorption lines and the Balmer break strength, as both depend on viewing angle. Such a potential correlation has recently been reported from the grating sample of \citet{Matthee2026}. However, such a putative correlation could also be driven by the outflows persisting to larger radii, which may clear a lower-column-density channel through the surrounding cold neutral reservoir. This is beyond the scope of the current \sirocco modelling, but could potentially affect the reservoir's spectral imprints, including its associated dust column.

\subsection{Modified blackbodies approximating \sirocco}
\begin{figure*}
    \vspace{-0.1cm}
    \includegraphics[angle=0,width=\textwidth]{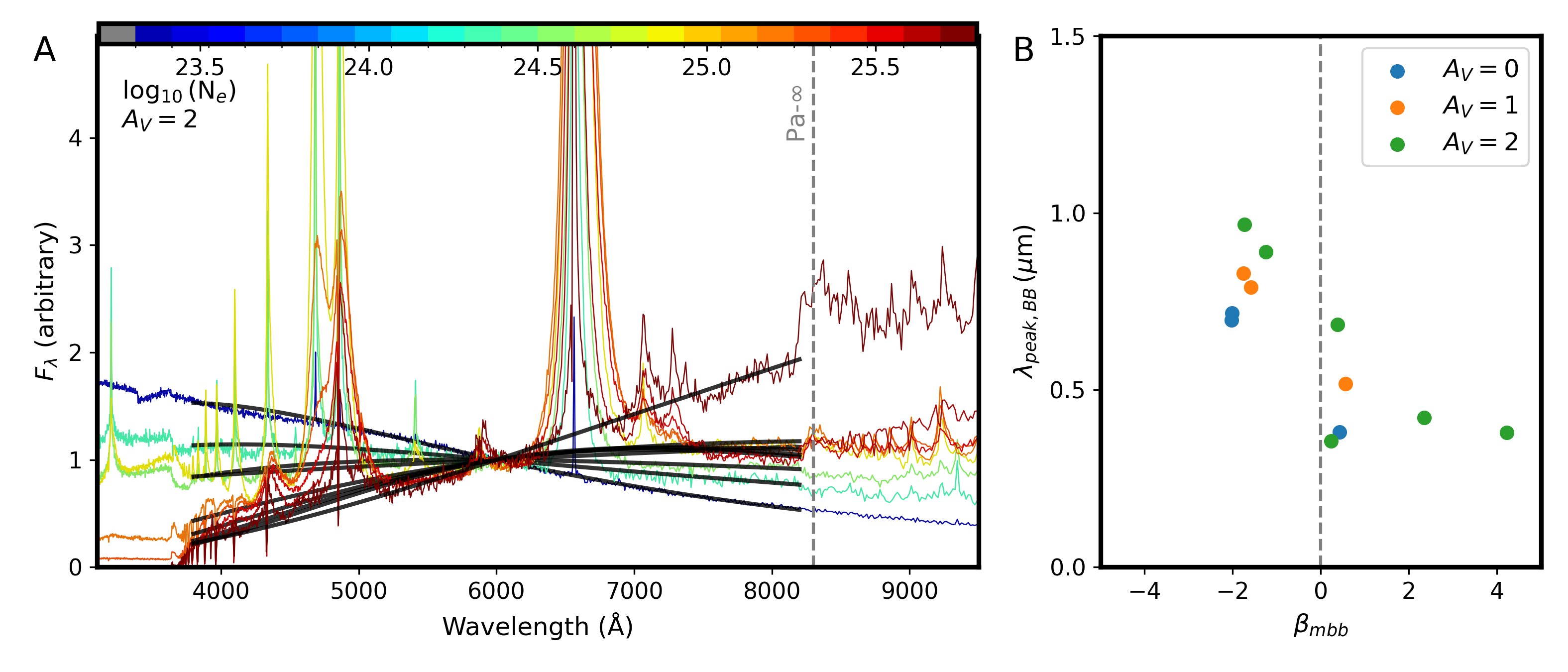}
    \vspace{-0.8cm}
\caption{{Blackbody fits to \sirocco models.} \emph{A}: Sequence of \sirocco models with varying column densities fit with modified Planck functions. \emph{B}: Blackbody peak wavelength as a function of the modification to spectral slope, $\beta_{\rm mbb}$. The \sirocco models reproduce a similar range of spectral peak wavelengths and $\beta_{\rm mbb}$ values and apparent correlations between these as the LRD sample in \citet{deGraaff2025b}, though the spectra are not blackbodies.}\label{fig:mbb_beta}
\end{figure*}

We fit modified blackbodies to the \sirocco-model spectra in Fig.~\ref{fig:exponential-sequence} including a modest level of extinction. We can fit the continua of the model spectra well with modified blackbodies, as found in observations \citep{deGraaff2025b}. Similar to these observational findings, the model spectra can be both narrower and broader than blackbodies (i.e.\ with blackbody $\beta_{\rm mbb}$ values between $-4$ and 3; see Fig.~\ref{fig:mbb_beta}). We emphasise that the \sirocco models are not optically thick to continuum absorption opacity and thus do not reprocess to the maximum entropy distribution; however, they still produce blackbody-adjacent continua. The red tail of emission in \sirocco is generally shallower than the Rayleigh-Jeans tail, while dust in a surrounding colder gas cloud or in the host galaxy ISM can redden the spectrum further, providing a wider range of rest-frame near-infrared spectral slopes. Opacity due to H$^-$ is not currently included in \sirocco and may produce more blackbody-like spectral shapes; however, given how closely our models currently match the data, additional continuum opacity is unlikely to change our major conclusions.

\subsection{Dust-to-gas ratio}\label{app:dust}
In the main text, we also note that we can infer a dust-to-gas ratio in the ionised gas cocoon. Since we can see that the sequence of observed widths is not monotonic with wavelength, this tells us that dust absorption does not dominate (i.e.\ it is not generally true in the observations that $v_\mathrm{e,H\beta}<v_\mathrm{e,H\alpha}<v_\mathrm{e,Pa\beta}$). This implies the dust absorption opacity at \hb is much lower than the continuum bound-free opacity at \ha within the gas responsible for scattering. This permits a direct observational limit on the dust-to-gas ratio within the cocoon:
\begin{equation}
\begin{aligned}
    \frac{M_\mathrm{dust}}{M_\mathrm{gas}}
    &\ll \frac{\kappa_\mathrm{gas,bf}(\mathrm{H}\alpha)}
              {\kappa_\mathrm{dust}(\mathrm{H}\beta)} 
    = \frac{n_3 \sigma_\mathrm{bf}(\mathrm{H}\alpha)}{\rho}
       \frac{1}{\kappa_\mathrm{dust}(\mathrm{H}\beta)} \\
    &= \frac{f_{n=3}\,\sigma_\mathrm{bf}(\mathrm{H}\alpha)}
             {\mu m_\mathrm{p}}
       \frac{1}{\kappa_\mathrm{dust}(\mathrm{H}\beta)} ,
\end{aligned}
\end{equation}
where $\rho$ is gas density, $n_3$ is the hydrogen density in $n=3$, $f_{n=3}$ is the fractional population, $m_p$ is the proton mass and $\sigma_\mathrm{bf}(\mathrm{H}\alpha)$ is the bound-free opacity evaluated at the \ha wavelength. The dust-absorption opacity can be estimated from the Milky Way dust-to-gas ratio ($\frac{M_\mathrm{gas}}{M_\mathrm{dust}}\approx100$) and the extinction optical depth per gas mass ($\tau_{\mathrm{H}\beta}/(N_\mathrm{H}\mu m_\mathrm{p})\approx100 \,\mathrm{cm^{2}/g}$; \citealt{Weingartner2001}): 
\begin{equation}
    \kappa_\mathrm{dust}(\mathrm{H}\beta) = \kappa_\mathrm{gas,dust}(\mathrm{H}\beta) \frac{M_\mathrm{gas}}{M_\mathrm{dust}} = \frac{\tau_{\mathrm{H}\beta}}{\mu m_\mathrm{p} N_\mathrm{H}} \frac{M_\mathrm{gas}}{M_\mathrm{dust}} = 10^{4} \mathrm{\,cm^2/g}
.\end{equation}
The implied dust-to-gas mass ratio in LRD cocoons is $\frac{M_\mathrm{dust}}{M_\mathrm{gas}}\ll 10^{-7}$ for our \sirocco $f_{n=3}$ population and similar stringent limits are produced in the LTE level-population limit with $T=9{,}000\,\mathrm{K}$. LRDs are often reddened, likely at least partly due to dust, but this measurement shows the dust occurs outside the region probed by the dense hot gas cocoon itself. Any dust responsible for the reddening must instead lie at larger radii, for example in the surrounding colder cocoon and/or in the host galaxy. The typical UV-slope of LRDs with $\beta_{UV} = -1.4\pm0.5$ \citep{deGraaff2025b} is consistent with moderately reddened stellar populations, e.g.\ $A_V\sim0.8$ \citep{PerezGonzalez2026}, so the UV continuum slope alone does not strongly constrain the location of that dust.

Having probed the importance of bound-free opacity from $n=3$, it is worth noting that this also implies that Paschen jumps should be observed. To our knowledge, this has not previously been observationally claimed and it is often difficult to assess given the high-redshift nature of LRDs and the observational requirement for continuum flux redward of rest-frame 1\,\micron. However, we note that fitting the continuum redward and blueward of the Paschen break for lower-redshift LRDs can provide distinct continua.
Finally, Balmer jumps can also be produced 
but the Paschen jump is more robust in the models and more distinguishable observationally, given the effects of dust reddening and blue contamination from stellar emission. While Balmer jumps are by definition not associated with the LRD population as currently defined, they have been a feature seen in a subset of high-redshift galaxies \citep{Cameron2024,Katz2025}. Our work suggests that these objects may also be LRD-like objects. We defer more detailed investigation of these findings to later work.

\subsection{Stability of the cocoon structure and reservoir mass}\label{app:stability}

The stability of a structure with a smaller envelope above a much larger mass has long been in question and arises in quasi-star and black hole envelope models \citep{Begelman2025,Kido2025,Sun2026}, while corresponding general issues in feeding SMBHs have been a long-standing issue \citep{Hopkins2010,Combes2023}. What we're proposing here is not a static structure, but a dynamical replenishment of the cocoon from a much larger reservoir. 
For an illustrative density profile $\rho(r)=10^{-14}\,{\rm g\,cm^{-3}} \left(\frac{r}{10^{16} {\rm cm}}\right)^{-2}$ and extrapolating to giant molecular cloud scales at $\sim30{-}100$\,pc, the total reservoir gas mass would be $M_{\rm res}\!\sim\!10^6\,\rm M_{\odot}$. Such an extrapolation is highly uncertain, but it illustrates that a much larger, mostly neutral reservoir could in principle be important for the longer term dynamical stability of the system, and indeed, is almost certainly necessary to replenish the relatively small inner ionized cocoon.
The inner, ionized cocoon should then be considered as only the innermost part of the larger cloud and can be thought of not as a smooth structure that would very rapidly collapse to a disk, but as a dynamical and radiatively self-regulated system, where feedback close to the Eddington luminosity supports a very mild, baryon-loaded fountain-like outflow that barely reaches escape velocity, and where much of the gas is recycled into an inflow that is perhaps not well-ordered and smooth and may well be clumpy
and filamentary. 
While a thin disk in the innermost part of the cocoon is plausible, a disk that mediates the accretion and provides the driving radiation, the rest of the cocoon outside the disk has a very large covering factor and reprocesses the accretion disk radiation into recombination emission, and provides the scattering medium and, in the cooler outer parts, the Balmer absorption. The self-regulation of the accretion may diminish long-term variability, while (as discussed elsewhere), the diffusion of radiation through the cocoon partially suppresses short timescale variability. The matter in the cocoon is depleted on a relatively short timescale and is continuously supplied from the cooler outer reservoir.  The cocoon causes most of the column density, while the reservoir provides most of the mass.  In this way, the dominant timescale is likely either the reservoir depletion timescale or the feedback disruption timescale.

\section{Asymmetry in the exponential wings}\label{app:asymmetry}

\begin{figure*}
\vspace{-0.4cm}
\begin{center}
    \includegraphics[angle=0,width=0.9\textwidth]{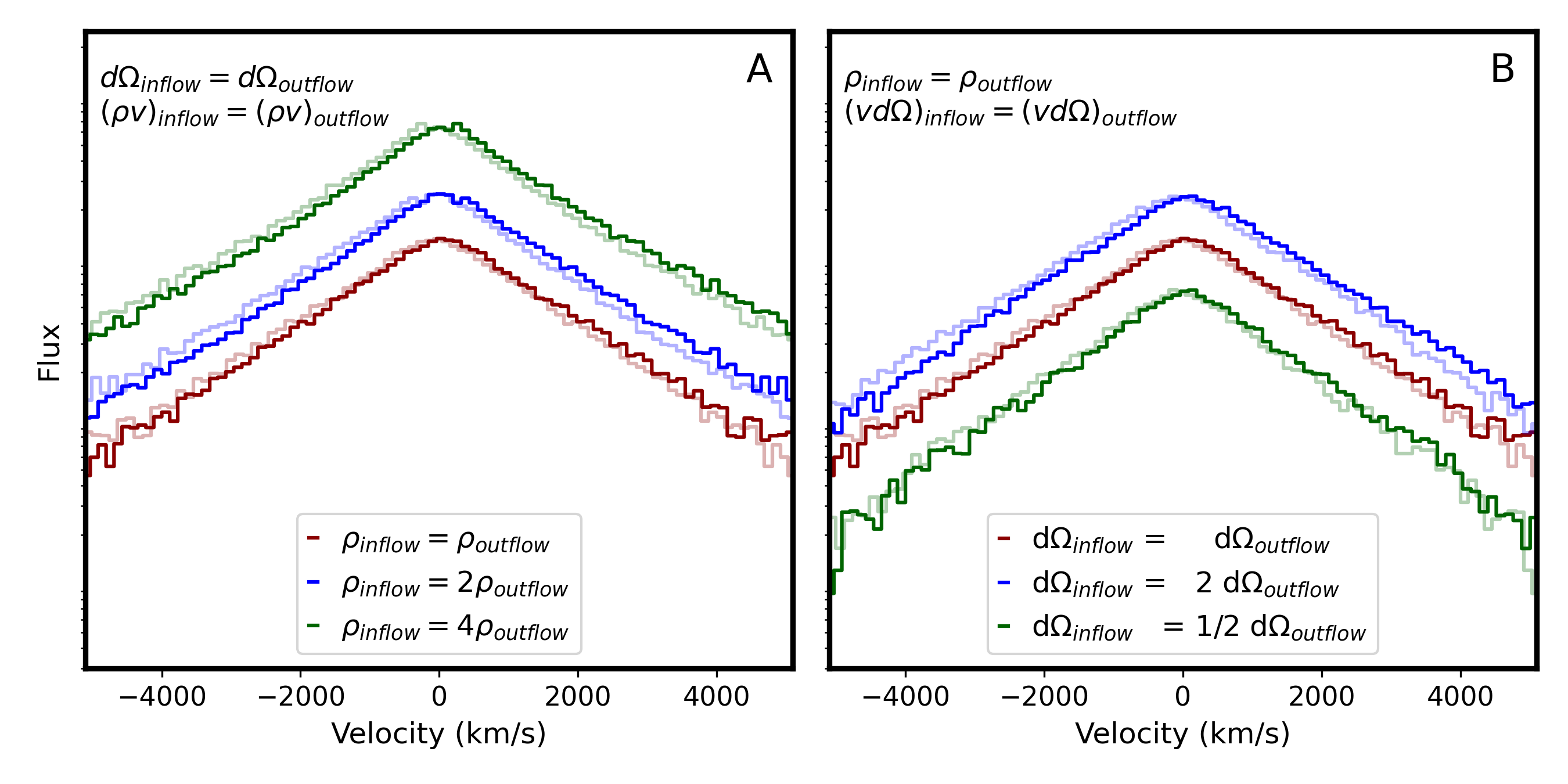}
\end{center}
\vspace{-0.8cm}
\caption{{Exponential line-shape with varying inflow/outflows}. Electron-scattering wings of \ha with parameters like Fig.~\ref{fig:viewing}, but with a variety of densities (\emph{left}), solid angles (\emph{right}) and velocities (\emph{both}). Here we show the line-profile of scattered photons as well as the mirrored profile (in lighter shading), to emphasise the symmetry in these model spectra. 
The common constraint on all model spectra shown here is that the mass flux of polar and equatorial components are equal. These model spectra are all consistent with symmetry at the level of the observations (i.e.\ $\sim5\%$).
}
\label{fig:fine_tuned}
\end{figure*}

\begin{figure}
\vspace{-0.3cm}
\includegraphics[angle=0,width=\linewidth]{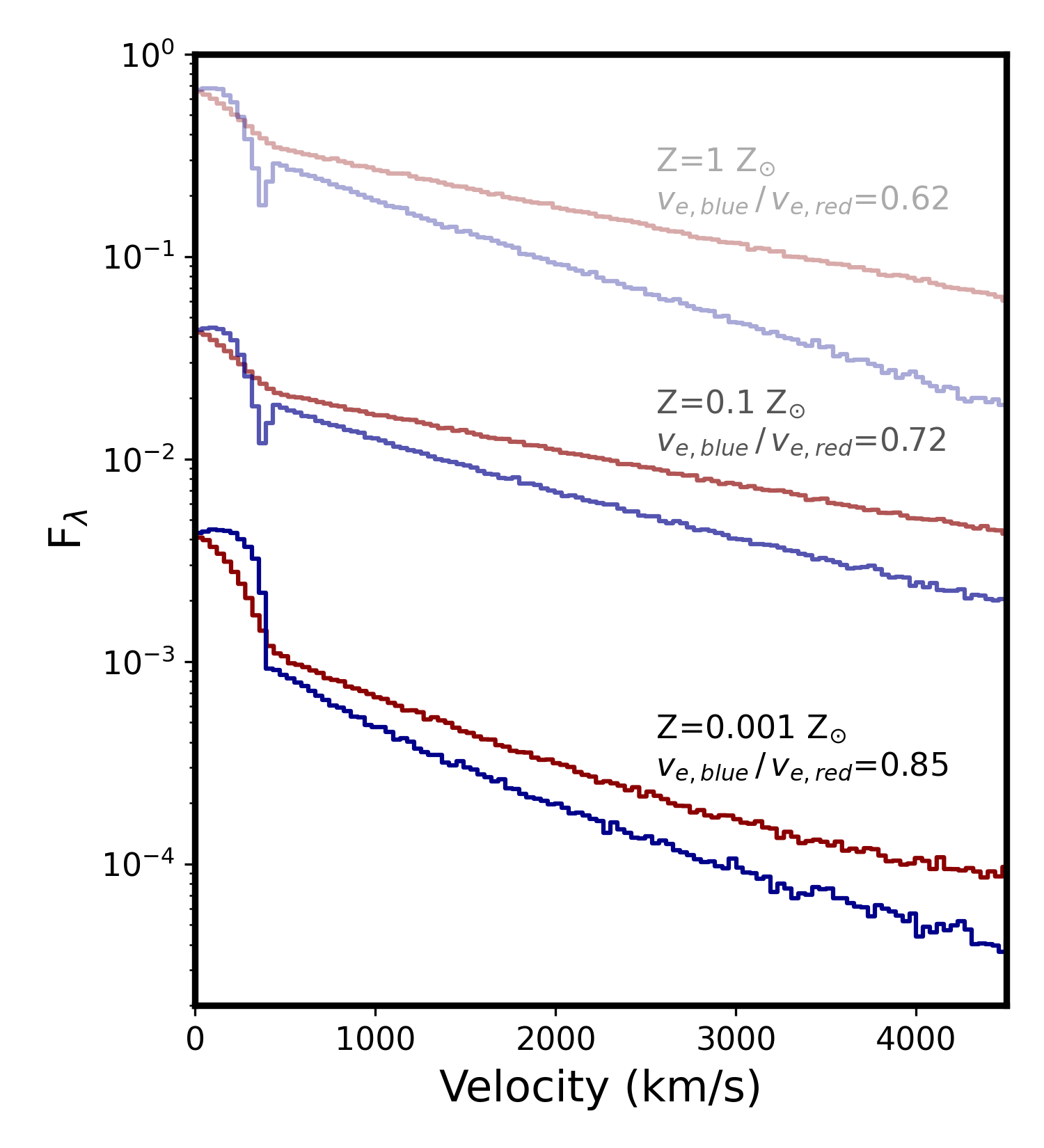}
\caption{{Effect of metallicity on the line shapes.} Here we show red/blue electron-scattered line shapes for a kinematically spherical electron-scattering medium at different metallicities. Metallicity affects the cooling, so that lower metallicities result in higher equilibrium temperatures and thus more symmetric profiles because the optical depth is lower for the same line width.}\label{fig:wind_metallicity}
\end{figure}

Here we explore the degree of line asymmetry produced by the electron-scattering medium. For velocities of a few hundred km\,s$^{-1}$, the asymmetry should be substantial (see Fig.~\ref{fig:wind_velocity}), which is in strong tension with the almost symmetrical broad lines observed (see Fig.~\ref{fig:sphericity}). Since the majority of observed LRDs are consistent with symmetric profiles, it is worth quantifying how fine-tuned this radiative-transfer condition is; this motivation also led us to search for observational candidates that are measurably asymmetric in the line shape, such as `Glimpse 17775' (Fig.~\ref{fig:Glimpse}).
We quantify here the deviations from exponential symmetry given the cocoon's inflow/outflow structure and find that near-exponential symmetry follows for comparable mass flux in outflows and inflows. We further consider alternate interpretations of the observed symmetry due to i) low metallicities, $Z\lesssim10^{-3}\,Z_{\odot}$, that reduce cooling and maintain higher electron temperatures or ii) multiple velocity components at different radii, which can produce scattering- and absorption-dominated features, respectively.

Blueshifted absorption features probe velocity outflows \textit{along the line-of-sight}, while the asymmetry of the exponentials measures \textit{the net kinematics} of the scattering gas as a whole. If the cocoon were a single monotonic outflow (i.e.\ a positive velocity divergence everywhere), an observable red--blue asymmetry would be unavoidable. By contrast, if inflows and outflows with comparable momentum coexist, their positive and negative divergences can partially cancel, yielding nearly symmetric wings. 
Simultaneous inflows and outflows would not be surprising in an accreting SMBH scenario of course, where the AGN luminosity arises from accretion, thus requiring an inflow, and feedback must produce an outflow.

\subsection{Deviations from symmetry}

The relative contributions of the inflowing and outflowing gas to the exponential wings scale with the number of scatterings (i.e.\ with the electron column density), while the degree of asymmetry is dictated by the bulk velocity. Consequently, comparable mass-flux inflows and outflows naturally produce symmetric wings. The setup illustrated in Fig.~\ref{fig:viewing} gives highly symmetric wings from all viewing angles when the inflow and outflow components have identical densities, absolute velocities, and covered solid angles. However, the inflow velocities could reasonably be expected to be much lower than the outflow velocities, as is often the case in accreting systems \citep{Yuan2014,Jorgensen2022}. This more general case is addressed in Fig.~\ref{fig:fine_tuned}, where we explore different velocities and densities (left panel) and solid angles (right panel) under the constraint that the net momentum in inflows and outflows is identical. The exponential profiles produced by \sirocco with comparable momentum in the inflows and outflows (see Fig.~\ref{fig:fine_tuned}) are symmetric to within $\sim5\%$, similar to the observational constraints of the typical LRD population, despite differences in the velocity or density. Conversely, non--mass-flux conserving flows easily produce asymmetries at levels $\gtrsim 10\%$. From these models, and the observed near-symmetry of the broad lines in most cases, we can infer that the mass fluxes in LRD-like objects must be similar in the inflows and outflows. 

If the inflow becomes sufficiently dense/optically thick then photons penetrate these regions less, and the resulting scattering profile becomes increasingly dominated by the outflow region. In particular, electron-scattering wings in TDEs are dominated by the outflows and therefore do not maintain symmetric exponential-wings \citep{Roth2018,Parkinson2022}. 
Thus, for sufficiently high SNR data and/or for extremely dense cocoons, the differences between inflows and outflows should cause asymmetries to emerge. Indeed, asymmetries are most significant in the highest SNR objects in the sample (Glimpse 17775, Rosetta Stone, etc.). While the \sirocco model spectra highlight empirical deviations from spherical kinematic symmetry, they do not uniquely constrain more detailed departures from spherical symmetry, such as azimuthal asymmetry, provided these still satisfy the momentum-weighted constraints of the scattering electron gas and reproduce the strong line-of-sight absorption.

\subsection{Alternative origins of symmetry}

Alternate interpretations of the observed highly symmetric exponentials could be due to (i) radially separated kinematic components, where the observed absorption line gas does not have the same kinematics as the scattering gas, (ii) large convective cells with inflowing and outflowing gas parcels or (iii) extremely high temperature gas, which would wash out kinematic information. 
Regarding argument (i), separate velocity components at different radii can fit many observed line profiles. However, the electron-scattering and Balmer-absorption regions occur on similar spatial scales, so it is difficult for radiative acceleration to produce an order-of-magnitude increase in the characteristic velocity between them. 
Large velocity gradients could arise if the accretion process is stochastic. In that case the layers producing scattering and absorption could trace different phases of activity, respectively a near static phase for the electron-scattering gas and an outflowing phase for absorption. However, this would require fine-tuning across the LRD population, since the electron-scattering layer is always observed in the near stationary state. 
Regarding (ii), convective cells with several simultaneous inflowing and outflowing radial parcels can produce symmetric electron-scattering wings. Such convective models can also be assessed within \sirocco, but as seen in Fig.~\ref{fig:convective_turb} (upper panels) these are generally associated with correspondingly more complex and centred absorption profiles than are typically observed, whereas observationally, the absorption is usually dominated by a relatively simple blue- or redshifted component.  
The fiducial \sirocco model, with a single large-scale inflow or outflow component, can produce deviations from perfectly symmetric exponential profiles. These asymmetries are increasingly diluted as the number of independent radial zones or velocity components increases. The middle panels of Fig.~\ref{fig:convective_turb} show the opposite limiting case: a turbulent velocity field with no large-scale velocity correlation. In this case, however, large-scale coherent inflow or outflow is still required to reproduce both the exponential line asymmetry and the observed red/blueshifted absorption features, as illustrated in the lower panels of Fig.~\ref{fig:convective_turb}. Thus, \sirocco models can span a continuum of velocity structures, ranging from coherent large-scale inflow/outflow to random local velocity perturbations, or combinations of the two. At present, the addition of local turbulent perturbations is not clearly required by the data. Nevertheless, such perturbations remain physically interesting because they can bring photons back into resonance at different locations within the cocoon, thereby reducing the effective diffusion length and increasing the equivalent width of absorption. However, the existence of significant line-of-sight absorption where the majority of the broad-line photons at some wavelengths get absorbed suggests that moving gas parcels should extend very roughly to the spatial scales of the cocoon. 

Regarding (iii), the high-temperature hypothesis is testable by asking whether the cocoon can sustain the required temperatures at LRD luminosities and plausible metallicities. This can be achieved in \sirocco with exceptionally high luminosities $L\gtrsim5\times10^{45}\,\mathrm{erg\,s}^{-1}$ and very low metallicities $Z\lesssim0.01\,Z_{\odot}$. However, across the LRD population with luminosities $L_\mathrm{bol}\sim10^{43}{-}10^{45}\,\mathrm{erg\,s}^{-1}$\citep{Greene2025}, highly symmetric exponentials cannot be achieved at any metallicity (see Fig.~\ref{fig:wind_metallicity}). Moreover, for a given observed wing width, higher $T_e$ implies a smaller required Thomson optical depth; in turn, photons should predominantly escape without scattering, which is generally inconsistent with the LRD population. So we can rule out this explanation.







\end{appendix}

\end{document}